\documentclass[12pt]{iopart}
\usepackage{graphicx,epsf,subfigure}
\usepackage{subeqnarray}
\usepackage{latexsym,amsbsy}

\begin{document}

\title{Bistable flows in precessing spheroids}

\author{D. C\'ebron}
\address{Universit\'e Grenoble Alpes, CNRS, ISTerre, Grenoble, France.}

\ead{david.cebron@ujf-grenoble.fr}

\begin{abstract}
Precession driven flows are found in any rotating container filled with liquid, when the rotation axis itself rotates about a secondary axis that is fixed in an inertial frame of reference. Because of its relevance for planetary fluid layers, many works consider spheroidal containers, where the uniform vorticity component of the bulk flow is reliably given by the well-known equations obtained by Busse in 1968. So far however, no analytical result on the solutions is available. Moreover, the cases where multiple flows can coexist have not been investigated in details since their discovery by Noir et al. (2003).  In this work, we aim at deriving analytical results on the solutions, aiming in particular at, first estimating the ranges of parameters where multiple solutions exist, and second studying quantitatively their stability. Using the models recently proposed by Noir \& C\'ebron (2013), which are more generic in the inviscid limit than the equations of Busse, we analytically describe these solutions, their conditions of existence, and their stability in a systematic manner. We then successfully compare these analytical results with the theory of Busse (1968). Dynamical model equations are finally proposed to investigate the stability of the solutions, which allows to describe the bifurcation of the unstable flow solution. We also report for the first time the possibility that time-dependent multiple flows can coexist in precessing triaxial ellipsoids. Numerical integrations of the algebraic and differential equations have been efficiently performed with the dedicated script FLIPPER (supplementary material).
\end{abstract}

\vspace{2pc}

\maketitle

\maketitle


\section{Introduction}

A rotating solid object is said to precess when its rotation axis itself rotates about a secondary axis that is fixed in an inertial frame of reference. In this work, we consider a precessing spheroid of fluid, in rotation around its symmetry axis. This geometry is indeed relevant for planetary fluid layers, such as the Earth liquid core \cite{greff1999core}, where precession-driven flows may participate in the dynamo mechanism generating their magnetic fields \cite{bullard1949magnetic,bondi1953dynamic,malkus1968precession}. These flows may also have an astrophysical relevance, for instance in neutron stars interiors where they can play a role in the observed precession of radio pulsars \cite{glampedakis2009}. Finally, precessing spheroids have been studied as turbulence generators, in particular when the angle between the container rotation axis and the precession axis is $90^{\circ}$ \cite{goto2007turbulence,Goto2014}. 

The first theoretical studies of this spheroidal geometry considered an inviscid fluid \cite{hough1895oscillations,sloudsky1895rotation,Poincare:1910p12351}. Assuming a uniform vorticity, they obtained a solution, called Poincar\'e flow, given by the sum of a solid body rotation and a potential flow. However, the Poincar\'e solution is modified by the apparition of boundary layers, and some strong internal shear layers are also created in the bulk of the
flow \cite{stewartson1963motion,Busse1968}. In 1968, Busse have taken into account these viscous effects as a correction to the inviscid flow in a spheroid, by
considering carefully the Ekman layer and its critical regions
(see also \cite{Busse1968,zhang2010fluid,Zhang2014}). Based on these works, \cite{cebron2010tilt} and \cite{Noir2013} have proposed models for the flow forced in precessing triaxial ellipsoids. Note that, beyond this correction approach, the complete viscous solution (including the fine description of all the flow viscous layers) has recently been obtained in the particular case of a weakly precessing spherical container \cite{kida2011steady}. 

When the precession forcing is large enough compared to viscous effects, instabilities can occur, destabilizing the Poincar\'e flow. First, the Ekman layers can be destabilized \cite{lorenzani2001fluid2} through standard Ekman layer instabilities \cite{lingwood1997absolute,faller1991instability}. In this case, the instability remains localized near the boundaries. Second, the whole Poincar\'e flow can be destabilized, leading to a volume turbulence: this is the
precessional instability \cite{malkus1968precession}. This small-scale intermittent flow confirm the possible relevance of precession for energy dissipation or magnetic field generation, and has thus motivated many studies. Early experimental attempts \cite{vanyo1991geodynamo,vanyo1995experiments} to confirm the theory of Busse \cite{Busse1968} did not give very good results \cite{pais2001precession}. Simulations have thus been performed in spherical containers \cite{tilgner1999magnetohydrodynamic, tilgner2001fluid}, spheres \cite{noir2001numerical}, and finally in spheroidal containers \cite{lorenzani2001fluid,lorenzani2003inertial}, allowing a validation of the theory of Busse \cite{Busse1968}. Experimental confirmation of the theory has then been obtained in spheroids \cite{Noir2003}, a work followed by many experimental studies involving spheres \cite{goto2007turbulence,kida2010super,boisson2012}, and spherical containers \cite{triana2012}.

Finally, the dynamo capability of precession driven flows has been demonstrated in spheres \cite{tilgner2005precession,tilgner2007kinematic}, spheroids \cite{wu2009dynamo} and cylinders \cite{nore2011nonlinear}, allowing the possibility of a precession driven dynamo in the liquid core of the Earth \cite{kerswell1996upper} or the Moon \cite{dwyerNature}.

In this work, we focus on the precession forced flow described by the system of algebraic equations obtained by Busse in 1968 \cite{Busse1968} for precessing spheroids. As shown by Noir and co-workers \cite{Noir2003}, these equations reliably describe the flow but can lead to multiple solutions for particular ranges of parameters . So far, these multiple solutions cases have not been investigated in details. Indeed, the study of Noir and co-workers \cite{Noir2003} is the only work considering these possible multiple solutions, and they only calculate the stability of the solutions in certain cases. In particular, the ranges of parameters allowing multiple solutions are not known (which is also partly due to the absence of any analytical result). No analysis of these equations have been performed to obtain rigorous constraints on the solutions, or to estimate the solutions and their stability.

We propose here to tackle analytically these issues in order to obtain analytical estimates and scaling laws, to compare our results to the exact solutions, and to investigate analytically the solutions stability.

In section \ref{sec:prob}, we introduce the problem considered in this work in a general framework. In section \ref{sec:multsolu}, we first present few multiple solutions cases (section \ref{sec:example}) and then, in section \ref{sec:dynmod}, we introduce recently proposed theoretical models \cite{Noir2013}, which extends the Busse equations into a dynamical framework. Relying on this new approach, theoretical investigations are tractable (section \ref{sec:analcal}), and the obtained analytical estimates are then compared with the predictions of the Busse model. The stability of the solutions are studied in section \ref{sec:stab}.

  \begin{figure}
  \begin{center}
    \begin{tabular}{ccc}
      \includegraphics[scale=0.47]{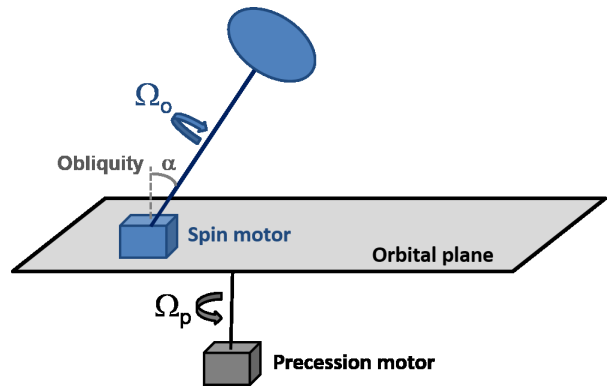}
     \end{tabular}
\caption{Sketch of the problem under consideration. We consider a container  filled with liquid and set in rotation at $\Omega_o$ along its symmetry axis $Oz$. The spheroid axis is tilted at the precession angle $\alpha$ and fixed on a rotating table, which  rotates at the precession rate $\boldsymbol{\Omega_p}$. In a planetary context, the fluid corresponds to a liquid core, the container to the mantle, and the rotating table plane to the orbital plane, i.e. the ecliptic for the Earth. }
    \label{fig:sketch1}             
  \end{center}
\end{figure}

\section{Mathematical description of the problem} \label{sec:prob}

We consider an incompressible homogeneous fluid of density $\rho$ and kinematic viscosity $\nu$ enclosed in a spheroidal cavity, rotating along its symmetry axis at $\boldsymbol{\Omega_o}=\Omega_o \boldsymbol{e_z}$ ($\boldsymbol{e_z}$ is a unit vector), and precessing at the angular velocity $\boldsymbol{\Omega_p}$, with a precession angle $\alpha$ , also called obliquity (see figure \ref{fig:sketch1}).
Noting $c$ the length of the spheroid symmetry axis, we use the length $a$ of the other principal axis as the length scale. Using $1/\Omega_o$ as a time scale, the problem is completely defined by four parameters: the oblateness $\eta_3=1-c/a$, or equivalently the flattening ellipticity $e=(c^2-a^2)/(c^2+a^2)$, the Ekman number $E=\nu/(\Omega_o a^2)$, the precession angle $\alpha$, and the Poincar\'e number $P=\Omega_p/\Omega_o$ in the literature. The dimensionless components of the precession vector are then naturally $P_x=P \sin \alpha$, $P_y=0$, and  $P_z=P \cos \alpha$ (the axisymmetry naturally allows the choice $P_y=0$ without any loss of generality). For planetary applications and turbulence generators, works of the literature typically consider moderate values of $e$, small Ekman number $E \ll 1$, and values of $\alpha$ are typically moderate for planetary cores ($\alpha \approx 23.4^{\circ} \approx  0.41\, \mathrm{rad} $ for the Earth), or large for turbulence generators studies ($\alpha=\pi/2$) \cite{goto2007turbulence,Goto2014}.

Considering the dimensionless fluid rotation rate  $\boldsymbol{\Omega_o}$, which is half the dimensionless vorticity, one can derive a system of equations governing the uniform vorticity bulk component of the flow (e.g. \cite{Noir2003,cebron2010tilt})
\begin{eqnarray}
   \Omega_x^2+\Omega_y^2 +\Omega_z^2 -\Omega_z &=& 0 , \label{eq:NoSpinUp}\\
    - P_z\ \Omega_y &=&  \eta_3\ \Omega_y \Omega_z + ( \lambda_r\ \Omega_x \Omega_z^{1/4}+\lambda_i\ \Omega_y \Omega_z^{-1/4})\sqrt{E} , \label{eq:Busse2} \\
   P_x\ \Omega_y  &=&   -\lambda_r\ \Omega_z^{1/4}\
   (1-\Omega_z)\ \sqrt{E}  \label{eq:Busse3} ,
\end{eqnarray}
which are exactly equations (20)-(22) of \cite{Noir2003}, or the equations (21)-(23) of \cite{cebron2010tilt} in the particular case of a spheroid ($\eta_2=0$ in their notations, and $P_y=0$). As shown by \cite{Noir2003}, this system of equations is equivalent to the well-known implicit expression (3.19) of Busse  \cite{Busse1968}. Equation (\ref{eq:NoSpinUp}) is the so-called no spin-up condition (see the solvability condition 3.14 of \cite{Busse1968}, or equation 12 of \cite{Noir2003}) given that it imposes $(\boldsymbol{\Omega}-\boldsymbol{e_z}) \cdot \boldsymbol{\Omega}=0$, i.e. it forbids any differential rotation along $\boldsymbol{\Omega}$. Equations (\ref{eq:Busse2})-(\ref{eq:Busse3}) are simply obtained from a torque balance (see \cite{Noir2003,cebron2010tilt} for details).

In equations (\ref{eq:NoSpinUp})-(\ref{eq:Busse3}), we have noted $\lambda=\lambda_r+\textrm{i} \lambda_i$ the spin-over damping factor, with $\lambda \approx -2.62+ 0.259 \textrm{i} $ for the sphere. For $E \ll 1$, an exact expression of $\lambda$ has actually been obtained \cite{zhang2004inertial}, which undergoes viscous corrections for finite values of $E$ \cite{hollerbach1995oscillatory,noir2001numerical}. 


Even if equations (\ref{eq:NoSpinUp})-(\ref{eq:Busse3}) are obtained without any inner core, corrections have been proposed in the case $a=c$ to take an inner core into account. Using the dimensionless inner radius $r_i$, it has been proposed to simply modify $\lambda$ by the factor $(1+r_i^4)/(1-r_i^5)$ for a no-slip inner core \cite{hollerbach1995oscillatory}, and by $1/(1-r_i^5)$ for a stress-free inner core \cite{tilgner2001fluid}.

It is possible to give a geometrical interpretation of equations  (\ref{eq:NoSpinUp})-(\ref{eq:Busse3}), which allows to obtain analytical constraints on the solutions. Since these constraints are interesting, but do not allow to really constrain the range of parameters allowing multiple solutions, we give this geometrical interpretation and the associated constraints in \ref{sec:geom}.

\section{Multiple stationary solutions in precessing spheroids} \label{sec:multsolu}

As shown by Noir and co-workers \cite{Noir2003}, equations (\ref{eq:NoSpinUp})-(\ref{eq:Busse3}) can actually be recast in a unique one, namely equation (24) of \cite{Noir2003}, which is exactly the implicit solution (3.19) of  \cite{Busse1968}. Little algebra shows that this unique equation can be transformed into a lengthy polynomial of degree $14$ (which reduces to degree $8$ in the sphere) for the unknown $\Omega_z$. This polynomial feature guarantees us to obtain efficiently all the possible solutions. However, one shall be careful since all the polynomial roots are not necessarily solutions of equations (\ref{eq:NoSpinUp})-(\ref{eq:Busse3}). We thus systematically test a posteriori the obtained solutions by replacing them into equations (\ref{eq:NoSpinUp})-(\ref{eq:Busse3}).

  \begin{figure}
  \begin{center}
    \begin{tabular}{ccc}
      \subfigure[]{\includegraphics[scale=0.5]{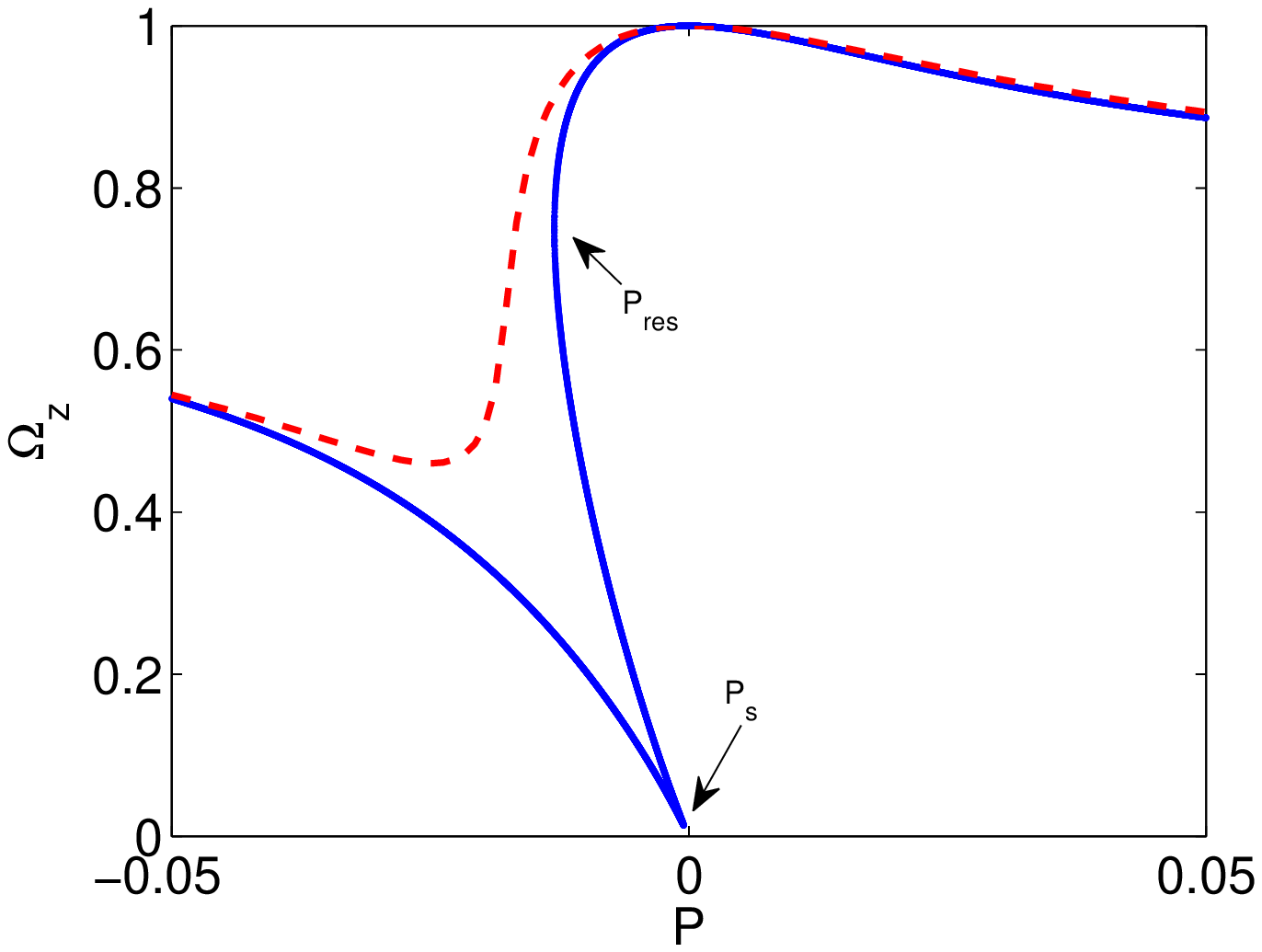}}
      \subfigure[]{\includegraphics[scale=0.5]{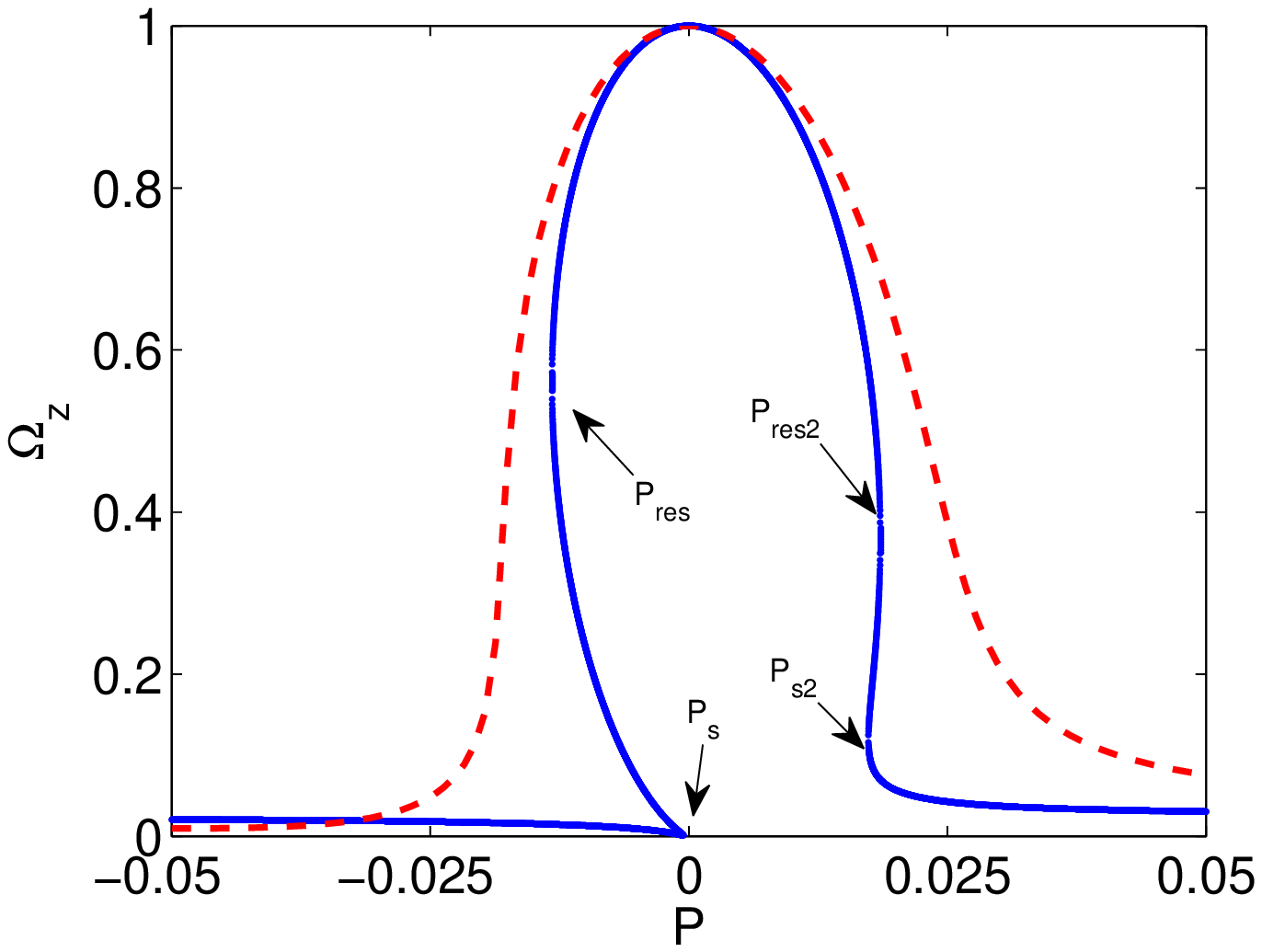}}
     \end{tabular}
\caption{Evolution of $\Omega_z=\Omega^2$ (see equation \ref{eq:NoSpinUp}), with the Poincar\'e number $P$ for $E=2.10^{-5}$ (dashed line) and $E=10^{-9}$ (solid line), with $c/a=0.97$, which gives $\lambda_r \approx -2.646$ and $\lambda_i \approx 0.306$ (inviscid values obtained from the formula of Zhang and co-workers \cite{zhang2004inertial}). (a) $\alpha=30^{\circ}$. (b) $\alpha=81^{\circ}$.}
    \label{fig:mult}             
  \end{center}
\end{figure}

\subsection{Examples of multiple solutions} \label{sec:example}
In figure \ref{fig:mult}a, solutions of  (\ref{eq:NoSpinUp})-(\ref{eq:Busse3}) are represented in function of the Poincar\'e number $P$, for a moderate precession angle of $\alpha=30^{\circ}$. This shows that, for large enough Ekman numbers (e.g. $E=2.10^{-5}$ in the figure), equations (\ref{eq:NoSpinUp})-(\ref{eq:Busse3}) lead to only one solution for each value of $P$. However,  when the Ekman number is smaller than a certain critical value $E_{max}$, certain values of $P$ can lead to multiple solutions (figure \ref{fig:mult}a). In this case, we can delineate three branches separated by a cusp point and a point where $\partial \Omega / \partial P=\infty$, and we note  $P_{s}$ and $P_{res}$ the respective associated Poincar\'e numbers (in figure \ref{fig:mult}a, $P_{res}<P_{s}$). Note that, according to Noir and co-workers \cite{Noir2003}, the branch between $P_{s}$ and $P_{res}$ is unstable, and cannot be physically realized. 

In figure \ref{fig:mult}b, we increase the precession angle to $\alpha=81^{\circ}$, leading to a quasi-symmetrical problem when $P$ is changed in $-P$ (naturally, the problem is exactly symmetrical $P \leftrightarrow -P$ for $\alpha=90^{\circ}$). This is reflected by the  vertical quasi-symmetry in figure \ref{fig:mult}. Considering the case $E=10^{-9}$, i.e. the case $E<E_{max}$, this figure shows that four singular points, delineating five branches, can exist when  $\alpha$ is larger than a certain value $\alpha_{lim}$. Starting from $P=-\infty$, we note the four singular points  as $P_{res}<P_{s} <P_{s2}<P_{res2} $. Note that this ordering is the opposite when $c/a>1$ (prolate spheroids), and that $P_{s2}$, and $P_{res2}$ only exist for large values of $\alpha$. We thus define non-ambiguously $P_{s}$ as the cusp point existing for any $\alpha$, $P_{s2}$ as the second cusp point appearing when $\alpha$ is large enough, and $P_{res}$, $P_{res2}$ as the points where $\partial \Omega / \partial P=\infty$. The point $P_{res}$ exist for any $\alpha$, whereas $P_{res2}$  only exists when $\alpha$ is large enough (in any case, $\eta_3 P_{res}<\eta_3 P_{s} <\eta_3 P_{s2}<\eta_3 P_{res2} $).

\begin{figure}
  \begin{center}
    \begin{tabular}{ccc}
      \subfigure[]{\includegraphics[scale=0.5]{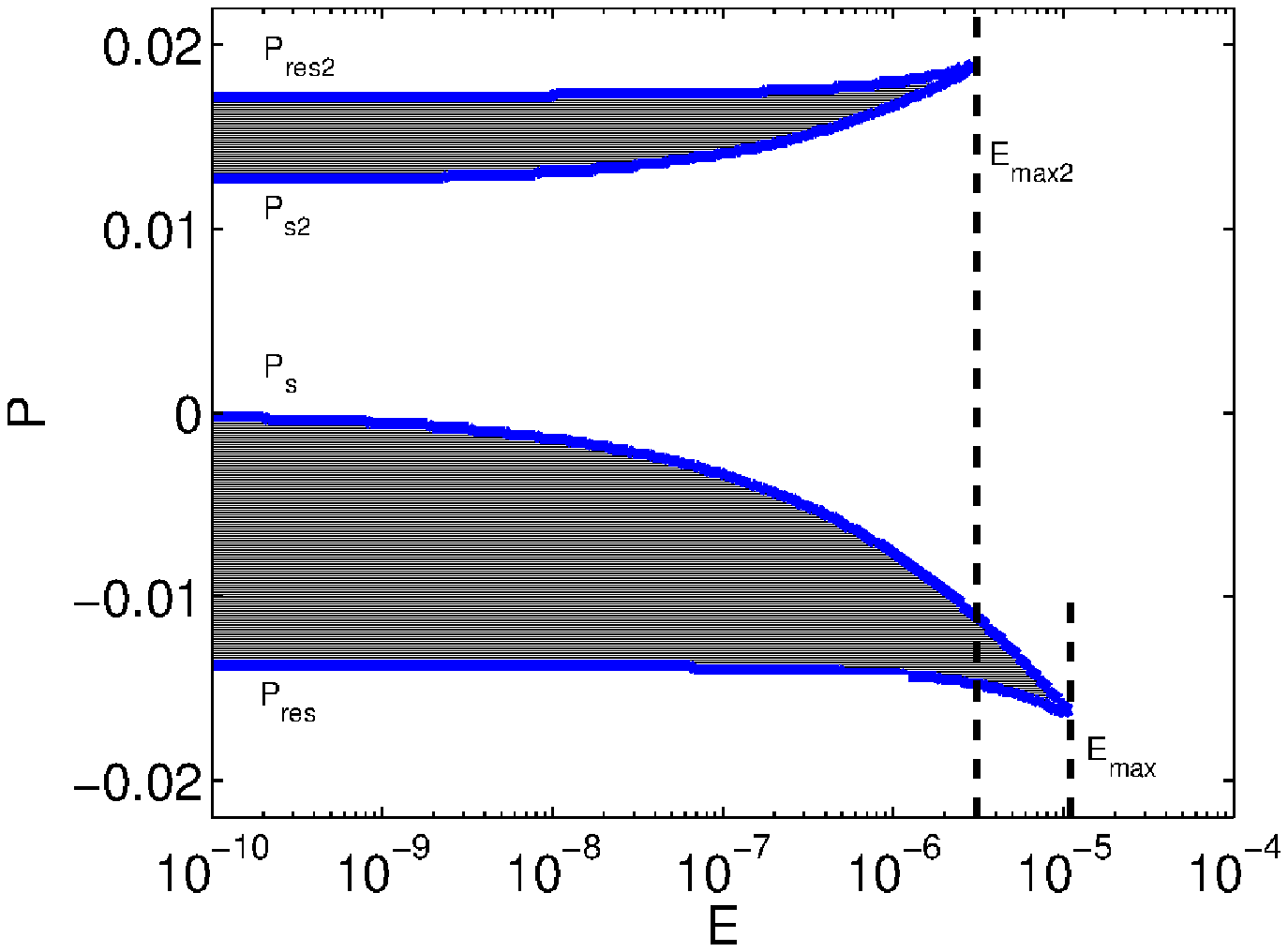}}
      \subfigure[]{\includegraphics[scale=0.5]{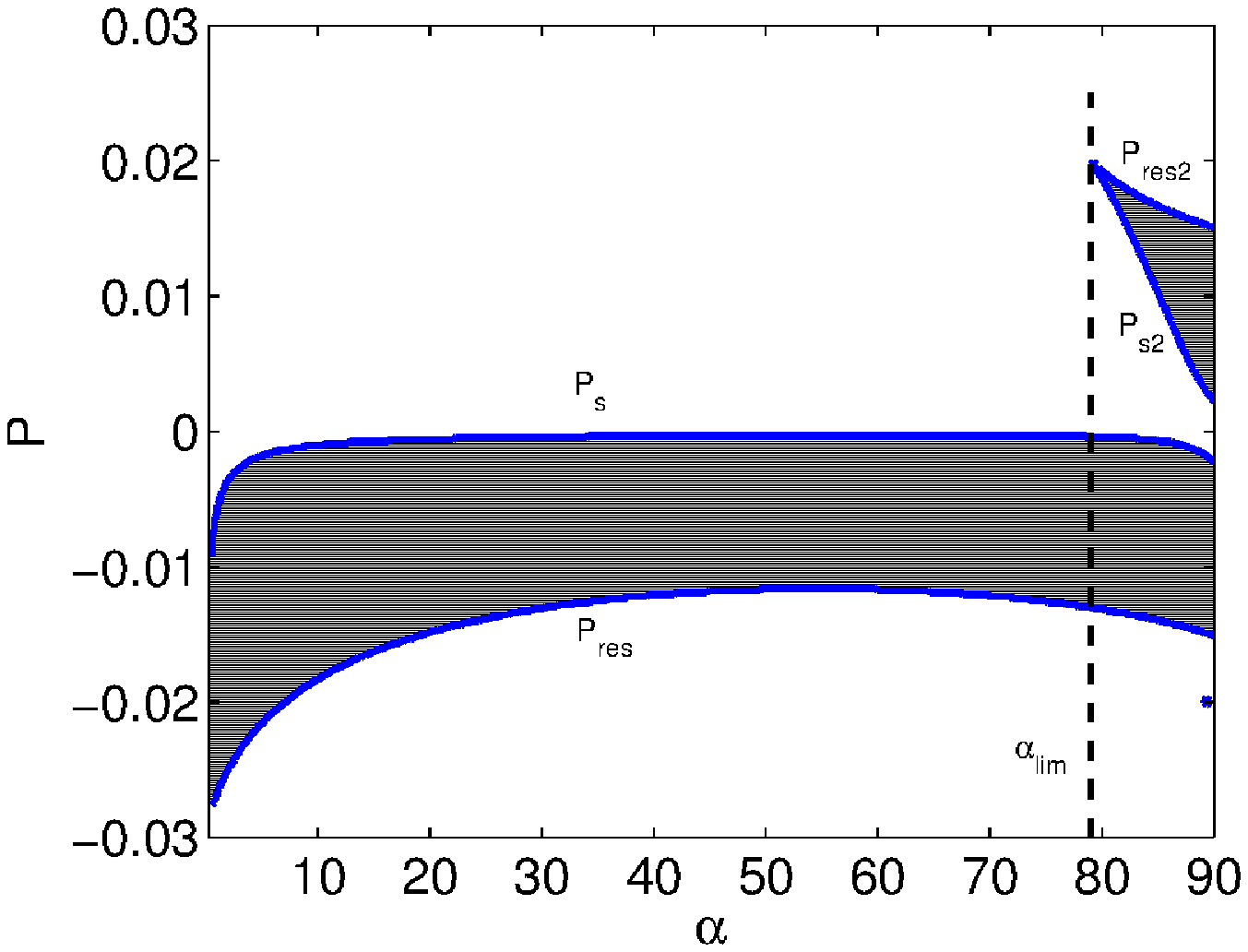}}
     \end{tabular}
\caption{Within the solid lines, i.e. for $P \in [P_{res};P_{s}]$ and $P \in [P_{s2};P_{res2}]$, equations (\ref{eq:NoSpinUp})-(\ref{eq:Busse3}) admit multiple solutions for $c/a=0.97$, which gives $\lambda_r \approx -2.646$ and $\lambda_i \approx 0.306$ (inviscid values obtained from the formula of Zhang and co-workers \cite{zhang2004inertial}). Then, the two figures show the ranges of $P$ leading to multiple solutions (hatched areas), (a) in function of $E$, for $\alpha=83.7^{\circ}$, and (b) in function of $\alpha$, for $E=10^{-9}$. We clearly see that $E_{max}$, obtained here around $E_{max} \approx 1.1\cdot 10^{-5}$, is actually given by $P_{res}=P_s$, whereas $\alpha_{lim}$, obtained here around $\alpha_{lim} \approx 79^{\circ}$, is actually given by $P_{res2}=P_{s2}$ ($P_{res2}=P_{s2}$ also defines $E_{max2}$, obtained here around $E_{max2} \approx 3.1\cdot 10^{-6}$).}
    \label{fig:mult2}             
  \end{center}
\end{figure}

In figure \ref{fig:mult2}, another point of view is proposed on these ranges of $P$ where equations (\ref{eq:NoSpinUp})-(\ref{eq:Busse3}) admit multiple solutions. In figure \ref{fig:mult2}a, we fix $\alpha=83.7^{\circ}>\alpha_{lim}$, which gives two zones with multiple solutions. The zone between $P_{res}$ and $P_s$ exists as soon as $E<E_{max}$, with $E_{max}$ given by $P_s=P_{res}$. The second zone, between $P_{res2}$ and $P_{s2}$ only exists at large precession angle ($\alpha>\alpha_{lim}$), and for $E<E_{max2}$, where $E_{max2}$ is given by $P_{s2}=P_{res2}$. In this figure, we clearly see the seven quantities we are interested in, i.e. $P_{res}$, $P_{s}$,  $P_{s2}$, $P_{res2} $, $E_{max}$, $E_{max2}$, and $\alpha_{lim}$, which are bounds of the multiple solutions areas. Our goal is thus to obtain estimates of these various quantities, in order to obtain analytical insights on these multiple solutions zoness.

\subsection{Busse stationary solutions seen as fixed points of a dynamical system} \label{sec:dynmod}
First, we did not manage to  obtain analytical results directly from equations (\ref{eq:NoSpinUp})-(\ref{eq:Busse3}). Second, within the mutiple solutions range of parameters, it would be interesting to test if these solutions can be realized experimentally, i.e. to study the stability in time of these solutions. To do so, equations (\ref{eq:NoSpinUp})-(\ref{eq:Busse3}) have to be obtained as fixed points of a dynamical model for $\boldsymbol{\Omega}$. Fortunately, these two issues have been recently tackled \cite{Noir2013}. Without any approximations, they show that $\boldsymbol{\Omega}$ is governed by (see equations A14-A 16 of \cite{Noir2013} for a spheroid)
\begin{eqnarray}
\frac{\partial \Omega_x}{\partial t}&=&P_z\Omega_{y}-e\left[P_z\Omega_{y}+\Omega_{y}\Omega_{z}\right]+ \mathcal{L} \boldsymbol{\Gamma}_{\nu}\cdot \boldsymbol{e}_x, \label{eq_axi_P_x}\\
%
\frac{\partial \Omega_y}{\partial t}&=&P_x\Omega_{z}-P_z\Omega_{x}+e\left[P_z\Omega_{x}+\Omega_{x}\Omega_{z}\right]+\mathcal{L} \boldsymbol{\Gamma}_{\nu}\cdot \boldsymbol{e}_y, \label{eq_axi_P_y}\\
%
\frac{\partial \Omega_z}{\partial t}&=&-P_x\Omega_{y}-e\, P_x\Omega_{y}+ \mathcal{L} \boldsymbol{\Gamma}_{\nu}\cdot \boldsymbol{e}_z, \label{eq_axi_P_z}
 \end{eqnarray}
with the viscous term $\mathcal{L} \boldsymbol{\Gamma}_{\nu}$. Contrary to equations (\ref{eq:NoSpinUp})-(\ref{eq:Busse3}), this reduced model does not assume small flattening for the inviscid part, and the results obtained in the inviscid limit $E=0$ will thus be valid for any oblateness. We will thus pay a particular attention to this limit, where this model gives more general and accurate results than equations (\ref{eq:NoSpinUp})-(\ref{eq:Busse3}).

So far, the expression of the viscous term $\mathcal{L} \boldsymbol{\Gamma}_{\nu}$  has not been obtained in the general case. However, in the limit $E \ll 1$, and if the angle between the fluid rotation vector and the container rotation vector is small, one can obtain an expression of $\mathcal{L} \boldsymbol{\Gamma}_{\nu}$ using the linear asymptotic expression of spin-up and of the spin-over mode (see the so-called generalized model, given by equation 2.28 of \cite{Noir2013}):
\begin{eqnarray}
 \mkern-36mu  \mathcal{L} \boldsymbol{\Gamma}_{\nu}= \sqrt{E\Omega} \left[ 
\frac{\lambda^r_{so}}{\Omega^2}\left( 
 \begin{array}{c}
      \Omega_x\Omega_z \\
      \Omega_y\Omega_z\\
      \Omega_z^2-\Omega^2\\
   \end{array}
   \right)+
 \frac{\lambda^i_{so}}{\Omega}\left(   
  \begin{array}{c}
     \Omega_y\\
     -\Omega_x\\
      0\\
   \end{array}
   \right)
   +
   \lambda_{sup} \frac{\Omega^2-\Omega_z}{\Omega^2} \left( 
 \begin{array}{c}
      \Omega_x\\
     \Omega_y\\
    \Omega_z\\
   \end{array}
   \right)
\right], \label{eq:GenVisc}
\end{eqnarray}
with
\begin{eqnarray}
\lambda_{sup}= - \frac{\sqrt{\pi^3/2}}{c\, \mathrm{\Gamma}(3/4)^2}\, \mathrm{F}\left([-1/4, 1/2],[3/4], 1-c^2 \right),
\end{eqnarray}
where $\mathrm{\Gamma}$ is simply the gamma function and $\mathrm{F}(n,d,z)$ is the usual generalized hypergeometric function, also known as the Barnes extended hypergeometric function (see respectively chap. 6 and 15 of \cite{Abramovitz}).

Naturally, equations (\ref{eq:NoSpinUp})-(\ref{eq:Busse3}), obtained in the very particular limit $P \ll 1$ and $e \ll 1$, are recovered as the fixed points of the dynamical model (\ref{eq_axi_P_x})-(\ref{eq_axi_P_z}) in this limit. For instance, taking (\ref{eq_axi_P_x})$\times \Omega_x+$ (\ref{eq_axi_P_y})$\times \Omega_y+$ (\ref{eq_axi_P_z})$\times \Omega_z$ yields
\begin{eqnarray}
(\boldsymbol{\Omega}-\boldsymbol{k})\cdot\boldsymbol{\Omega}=\frac{e P_x \Omega_y\Omega_z}{\lambda_{sup} \sqrt{E}}.
\end{eqnarray}
Then, in the limit $e P_x / \sqrt{E} \ll 1$, we recover the so-called no spin-up condition (\ref{eq:NoSpinUp}). This condition is thus not valid in general for a spheroid of arbitrary ellipticity. In this limit, which is the limit of validity of equations (\ref{eq:NoSpinUp})-(\ref{eq:Busse3}), the viscous term $\mathcal{L} \boldsymbol{\Gamma}_{\nu}$ reduces to
\begin{eqnarray}\label{viscousSO}
\mathcal{L} \boldsymbol{\Gamma}_{\nu}=(E\Omega)^{1/2}\left[ 
\lambda_r \left( 
 \begin{array}{ccc}
      \Omega_x \\
      \Omega_y\\
      \Omega_z-1\\
   \end{array}
   \right)+
 \frac{\lambda_i}{\Omega}\left(   
  \begin{array}{ccc}
     \Omega_y\\
     -\Omega_x\\
      0\\
   \end{array}
   \right)
   \right], \label{eqNoSUP}
\end{eqnarray}
by simply using the no spin-up equation $\Omega_z=\Omega^2$. 

Finally, (\ref{eq:NoSpinUp})-(\ref{eq:Busse3}) are simply the fixed points of the dynamical system given by equations (\ref{eq_axi_P_x})-(\ref{eq_axi_P_z}) and (\ref{eqNoSUP}), and the solutions stability can be calculated in this framework. However, we still need a simpler set of equations to perform tractable analytical calculations. To do so, we follow \cite{Noir2013}, who showed that the viscous term (\ref{eqNoSUP}) can be very well approximated by the so-called reduced form (equation 2.29 of their work \cite{Noir2013})
\begin{eqnarray}\label{addhocvisc}
\mathcal{L} \boldsymbol{\Gamma}_{\nu}= \lambda \sqrt{E} \left( 
 \begin{array}{c}
      \Omega_x\\
      \Omega_y\\
      \Omega_z-1\\
   \end{array}
\right),
\end{eqnarray}
which is linear in $\boldsymbol{\Omega}$. One can notice that the linear viscous terms of the reduced model does not include the coefficient $\lambda_i$, and thus neglect its influence (typically, a small viscous modification of the values of $P$ where the Poincar\'e flow undergoes a resonance, see \cite{Noir2013} for details). 

The analytical calculations presented in this work have been performed with the computer algebra system MAPLE. They have been compared with numerical solutions of algebraic and differential equations (e.g. equations \ref{eq:NoSpinUp}-\ref{eq:Busse3} and \ref{eq_axi_P_x}-\ref{eq_axi_P_z}, respectively) solved with the dedicated MATLAB script FLIPPER. This home-made script allows to solve efficiently all the equations described above, either by time-stepping or by directly looking for all the possible steady solutions. This script can also solve the system of equations proposed by \cite{cebron2010tilt} and \cite{Noir2013} for precessing triaxial ellipsoids, implementing the various viscous terms (equations (\ref{eq:GenVisc}, \ref{eqNoSUP}, and \ref{addhocvisc}), in each case. The script FLIPPER and its documentation are provided in supplementary materials.
  
  \subsection{Analytical estimates, using the so-called reduced model} \label{sec:analcal}
  The calculation details are given in \ref{sec:calc}, and we thus only report below the important analytical results and steps of this calculation.
  
Focusing on stationary solutions of equations (\ref{eq_axi_P_x})-(\ref{eq_axi_P_z}) with the viscous term (\ref{addhocvisc}), these equations can be recast in a unique polynomial of degree $3$ for the unknown $\Omega_z$, which allows tractable algebra investigations. For the sphere ($e=0$), this polynomial reduces to a linear polynomial, and the explicit solution for $\boldsymbol{\Omega}$ is then
\begin{eqnarray}
\Omega_x &=& \frac{P^2 \sin 2 \alpha}{2(P^2+\lambda^2\, E)}, \label{eq:sphereX} \\
\Omega_y &=& -\frac{P \sin (\alpha)\, \lambda_r \sqrt{E}}{P^2+\lambda^2\, E}, \\
\Omega_z &=& \frac{\lambda_r^2 E +P^2  \cos^2 \alpha}{P^2+\lambda^2\, E}. \label{eq:sphereZ}
\end{eqnarray}
For the sphere, there is thus always a unique solution for the reduced model. Note also that the solution (\ref{eq:sphereX})-(\ref{eq:sphereZ}) is interesting in the limit $E=0$. Indeed, calculating a uniform vorticity solution of the Euler equations for a precessing spheroid leads to the so-called Poincar\'e flow, which has a free parameter (as a consequence of the inviscid hypothesis). Among this class of solutions, the one usually chosen in the literature is defined arbitrarily by putting $\Omega_z=1$ \cite{tilgner2007rotational}. Here, equations (\ref{eq:sphereX})-(\ref{eq:sphereZ}) show that, for the sphere, the solution in the limit of vanishing viscosity is actually the Poincar\'e flow defined with $\Omega_z=\cos^2 \alpha$.
  
We consider below the case of a spheroid ($e \neq 0$). Little algebra shows that the multiple solutions boundaries $P_s$ and $P_{s2}$ (see figure \ref{fig:mult2}) can be described by a unique boundary $E_s=f(P)$, where $f(P)$ is an analytically known root of a polynomial of degree $3$. An expansion for $P \ll 1$ gives the following tractable expression at order $\mathcal{O}(P^5)$
\begin{eqnarray}
\lambda^2 E_s=\frac{1-e^2}{e} s_i^2 c_o P^3+\frac{1+e}{4e^2}s_i^2[4 c_o^2(1+e^2)+s_i^2(1+e)-8e c_o^2] P^4, \label{eq:Es}
\end{eqnarray}
where the branch $P_s$ is obtained for $\textrm{sign}(e) P>0$, and the branch $P_{s2}$ for $\textrm{sign}(e) P< \textrm{sign}(e)  P_{s2}^{inv}$. In the planetary relevant limit $e \ll1 $, equation (\ref{eq:Es}) gives
\begin{eqnarray}
P =\frac{\lambda^{2/3}}{s_i^{2/3} c_o^{1/3}}\, e^{1/3}\, E_s^{1/3}+\mathcal{O}(e^{7/3})\, \, \, \,  \, \, \,  \, \,  \textrm{for}  \, \, \alpha \neq \frac{\pi}{2}.
\end{eqnarray}
At the order $\mathcal{O}(P^5)$ of equation (\ref{eq:Es}), putting $E_s=0$ leads to two solutions for $P$. The first solution is $P=0$, corresponding to the inviscid limit $P_{s}^{inv}=0$ of $P_{s}$. The second solution corresponds to the inviscid limit $P_{s2}^{inv}$ of $P_{s2}$, given by
\begin{eqnarray}
P_{s2}^{inv}=\frac{4 e (e-1) c_o}{1+e+c_o^2 [3-e(9-4 e)]}. \label{Pk10}
\end{eqnarray}
Note that a compact accurate expansion of (\ref{eq:Es}) can be obtained at the order $\mathcal{O}(P^{20})$ for the particular case $\alpha=\pi/2$:
\begin{eqnarray}
\lambda^2 E_s= \zeta_1+\zeta_2+\frac{3}{2} \zeta_3+\frac{11}{4} \zeta_4+\frac{91}{16} \zeta_5 +\frac{51}{4} \zeta_6+\frac{969}{32}\zeta_7+\frac{4807}{64}\zeta_8, \label{eq:Es90}
\end{eqnarray}
with $\zeta_k=(1+e)^{1+k} e^{-2k} P^{2(k+1)}/4$. Note also that the leading order of equation (\ref{eq:Es}) vanishes for $\alpha=\pi/2$, which imposes to consider the next order. In the limit $e \ll1 $, equation (\ref{eq:Es}) then gives
\begin{eqnarray}
P =\sqrt{2 \lambda}\, e^{1/2} \, E_s^{1/4}+\mathcal{O}(e^{3/2}) \, \, \, \,  \, \,  \textrm{for} \,  \, \,  \, \, \alpha = \frac{\pi}{2}.
\end{eqnarray}

Since $P_{res}$ and $P_{res2}$ are actually weakly dependent on $E$ (see e.g. fig. \ref{fig:mult2}), it turns out that their inviscid limit values, noted respectively $P_{res}^{inv}$ and $P_{res2}^{inv}$, provide good estimates of these quantities. They are given by
\begin{eqnarray}
 \mkern-72mu   P_{res}^{inv} \approx \left|
 \begin{array}{c}
 \displaystyle
       \alpha \ll 1: \frac{e}{1-e} - \frac{3e[2(1+e)]^{1/3}  \alpha^{2/3}}{2(1-e)^{5/3}}+ \frac{7e[2(1-e^2)]^{2/3}\alpha^{4/3}}{4(1-e)^3}  +\mathcal{O}(\alpha^{2}) \\
       \\
      \displaystyle
     x=\pi/2-\alpha \ll 1: \frac{e}{2\sqrt{1+e}}- \frac{e(1-e)}{2(1+e)} x +\frac{e (3 e^2-5 e+4)}{4(1+e)^{3/2}} x^2+\mathcal{O}(x^3) \\
   \end{array}
\right. \label{eq:Pres} 
\end{eqnarray}
and
\begin{eqnarray}
P_{res2}^{inv} \approx -  \frac{e}{2\sqrt{1+e}}-\frac{e(1-e)}{2(1+e)} x - \frac{e (3 e^2-5 e+4)}{4(1+e)^{3/2}} x^2+\mathcal{O}(x^3), \label{eq:Pres2inv}
\end{eqnarray}
for  $x=\pi/2-\alpha \ll 1$. It is naturally satisfying that the leading order of $P_{res}^{inv}$ for $\alpha \ll 1$ recovers the usual linear inviscid resonance $P_r=e/(1-e)$ of the Poincar\'e flow \cite{Poincare:1910p12351,cebron2010tilt}. It is important to note that we provide here, for the first time, an analytical estimate of the higher-order corrections to the linear resonance $P_r$ of the Poincar\'e flow, i.e an estimate of the so-called non-linear resonance \cite{Noir2003}. It is also satisfying to retrieve the quasi-symmetry of the problem (with respect to $P=0$) in the expressions of $P_{res}$ and $P_{res2}$ for $x=\pi/2-\alpha \ll 1$.

  \begin{figure}
  \begin{center}
    \begin{tabular}{ccc}
      \subfigure[]{\includegraphics[scale=0.47]{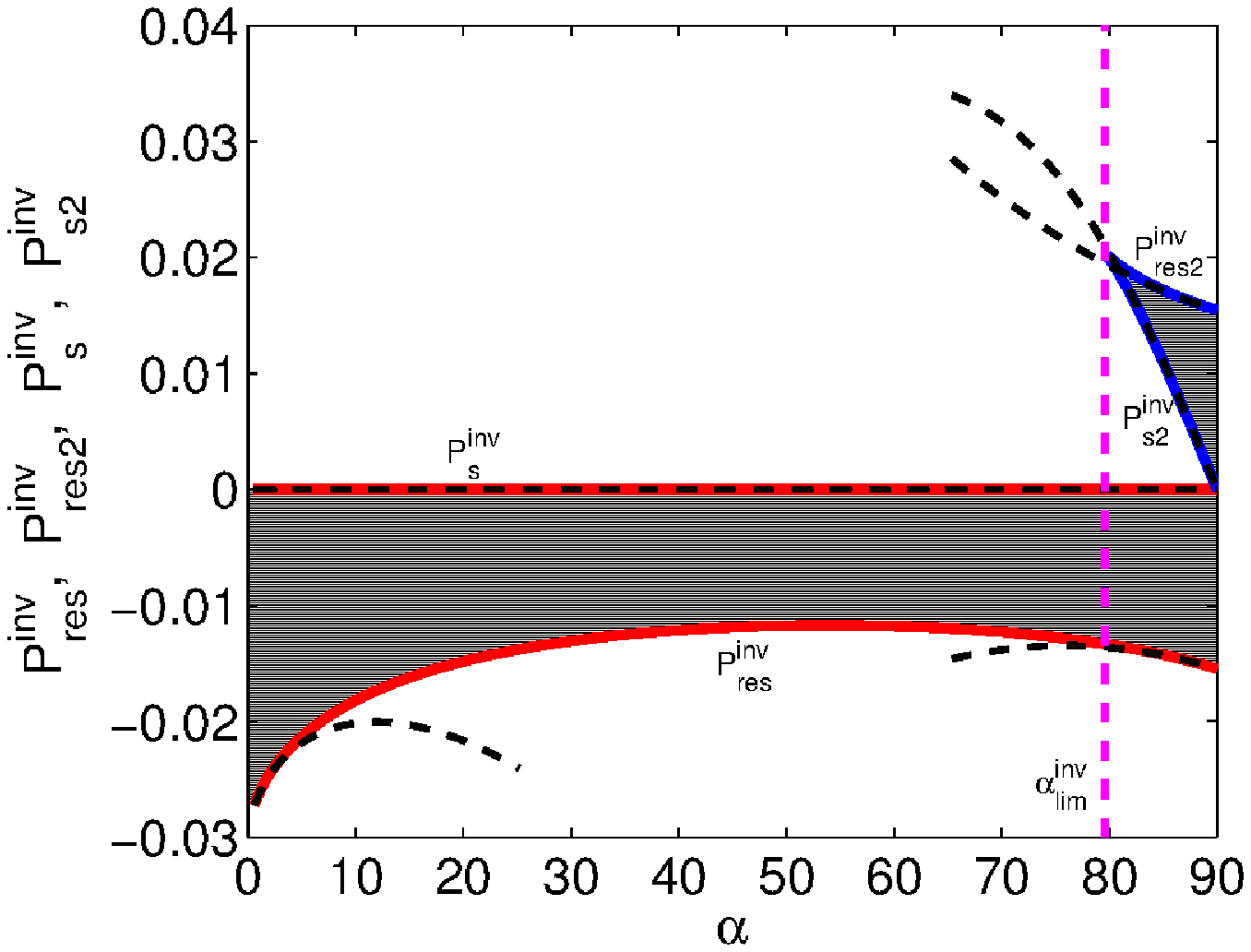}}
      \subfigure[]{\includegraphics[scale=0.47]{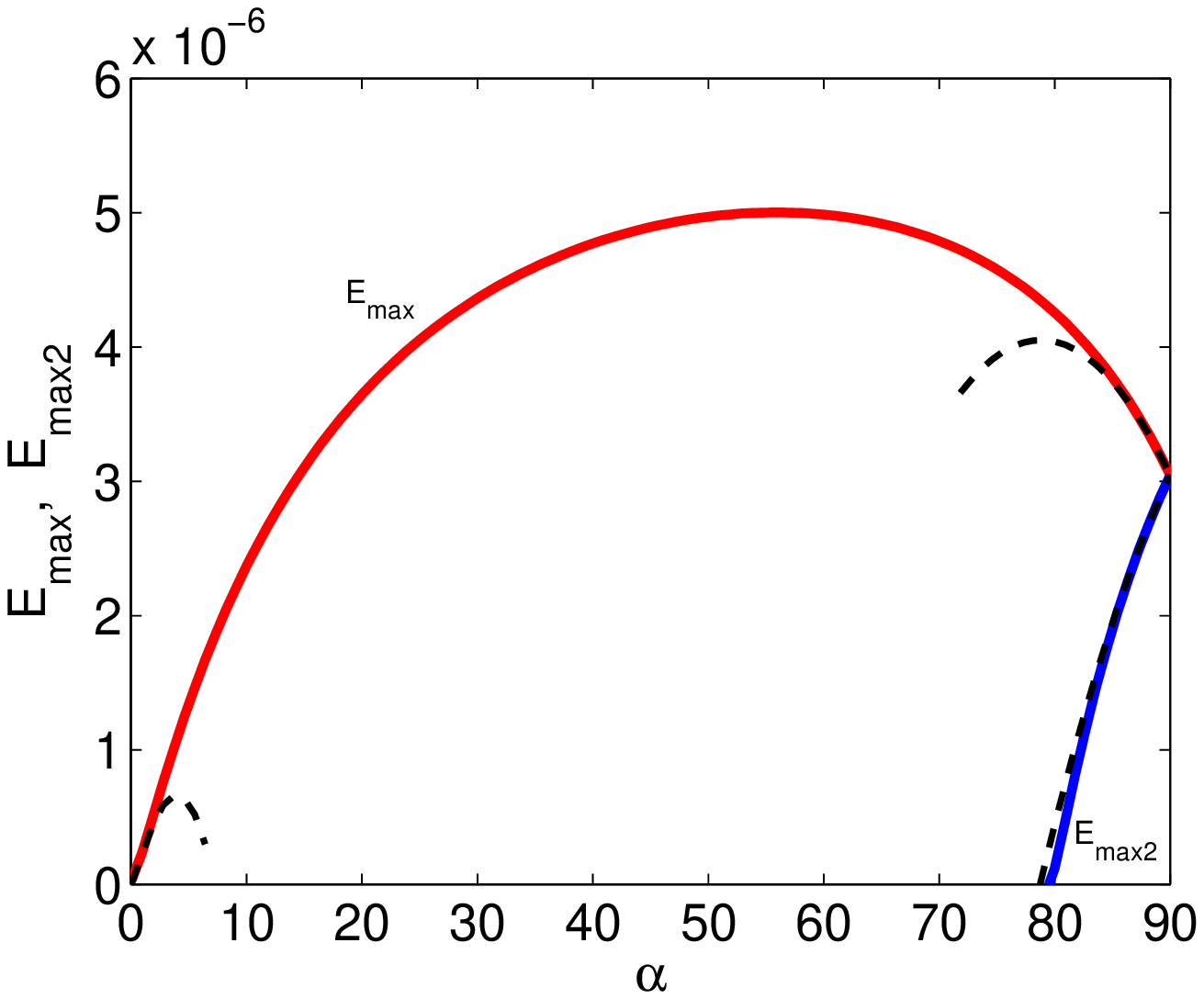}}
     \end{tabular}
\caption{We fix here $c/a=0.97$. (a) Thick solid lines correspond, from bottom to top, to $P_{res}^{inv}$, $P_{s}^{inv}$, $P_{s2}^{inv}$, $P_{res2}^{inv}$ (roots of $D_0$, see  \ref{sec:calc} for details), whereas the dashed lines are the associated expansions, respectively given by equation (\ref{eq:Pres}), $P_{s}^{inv}=0$,  equation (\ref{Pk10}), and equation (\ref{eq:Pres2inv}). The vertical dashed line is given by equation (\ref{eq:angleLIM}) for $ \alpha_{lim}^{inv}$. (b) Thick solid lines corresponds to $E_{max}$ and $E_{max2}$, respectively estimated by the two intersection points between $P_s$ and $P_{res}^{inv}$, as well as $P_{s2}$ and $P_{res2}^{inv}$ (i.e. the solutions of the system $\Delta=0,\, D_0=0$, see  \ref{sec:calc} for details). The dashed lines are the assiociated expansions, respectively given by equations (\ref{eq:EmaxA}) and (\ref{eq:Emax2A}). We have used here $\lambda_r=-2.62$.}
    \label{fig:alphEmax}             
  \end{center}
\end{figure}

Based on these estimates, one can calculate the critical values $\alpha_{lim}$ and $E_{max}$. The critical angle $\alpha_{lim}$ is well estimated by its inviscid limit $\alpha_{lim}^{inv}$, given by 
\begin{eqnarray}
| \alpha_{lim}^{inv} | = \textrm{arcos} \left( \sqrt{\frac{1+e}{27e^2-53e+28}} \right), \label{eq:angleLIM}
\end{eqnarray}
which is decreasing between $\pi/2$, reached for $e=-1$, and $0$, reached for $e=1$. Finally,  $E_{max}$ (resp. $E_{max2}$) is simply estimated by the intersection point between $P_{res}^{inv}$ and $P_s^{inv}$ (resp. $P_{res2}^{inv}$ and $P_{s2}^{inv}$). We thus obtain
\begin{eqnarray}
 \mkern-72mu \frac{\lambda^2 E_{max}}{e^2} \approx \left|
 \begin{array}{c}
 \displaystyle
       \alpha \ll 1: \frac{3(2\sqrt{3}-3) (\xi \alpha)^{4/3}}{2^{4/3}} -\frac{(13\sqrt{3}-18) (\xi \alpha)^2}{2}  +\mathcal{O}(\alpha^{8/3}) \\
       \\
      \displaystyle
     x=\pi/2-\alpha \ll 1:  \frac{-8+5 \phi}{4} +\frac{(5-3 \phi)x}{2 \xi} -\frac{(39-23 \phi) x^2}{10 \xi^2}+\mathcal{O}(x^3) \\
   \end{array}
\right.  \label{eq:EmaxA}
\end{eqnarray}
with $\xi=(1+e)^{1/2}/(1-e)$, and the golden ratio  $\phi=(1+\sqrt{5})/2$. Similarly, $E_{max2}$ is given by
\begin{eqnarray}
\frac{\lambda^2 E_{max2}}{e^2} \approx \frac{-8+5 \phi}{4} -\frac{(5-3 \phi)}{2 \xi} x -\frac{(39-23 \phi)}{10 \xi^2}  x^2+\mathcal{O}(x^3), \label{eq:Emax2A}
\end{eqnarray}
with  $x=\pi/2-\alpha \ll 1$. The problem symmetry for $\alpha=\pi/2$ is recovered in the expressions (\ref{eq:EmaxA}) and (\ref{eq:Emax2A}) since $E_{max}=E_{max2}$ in this case.

  \begin{figure}
  \begin{center}
    \begin{tabular}{ccc}
      \subfigure[]{\includegraphics[scale=0.47]{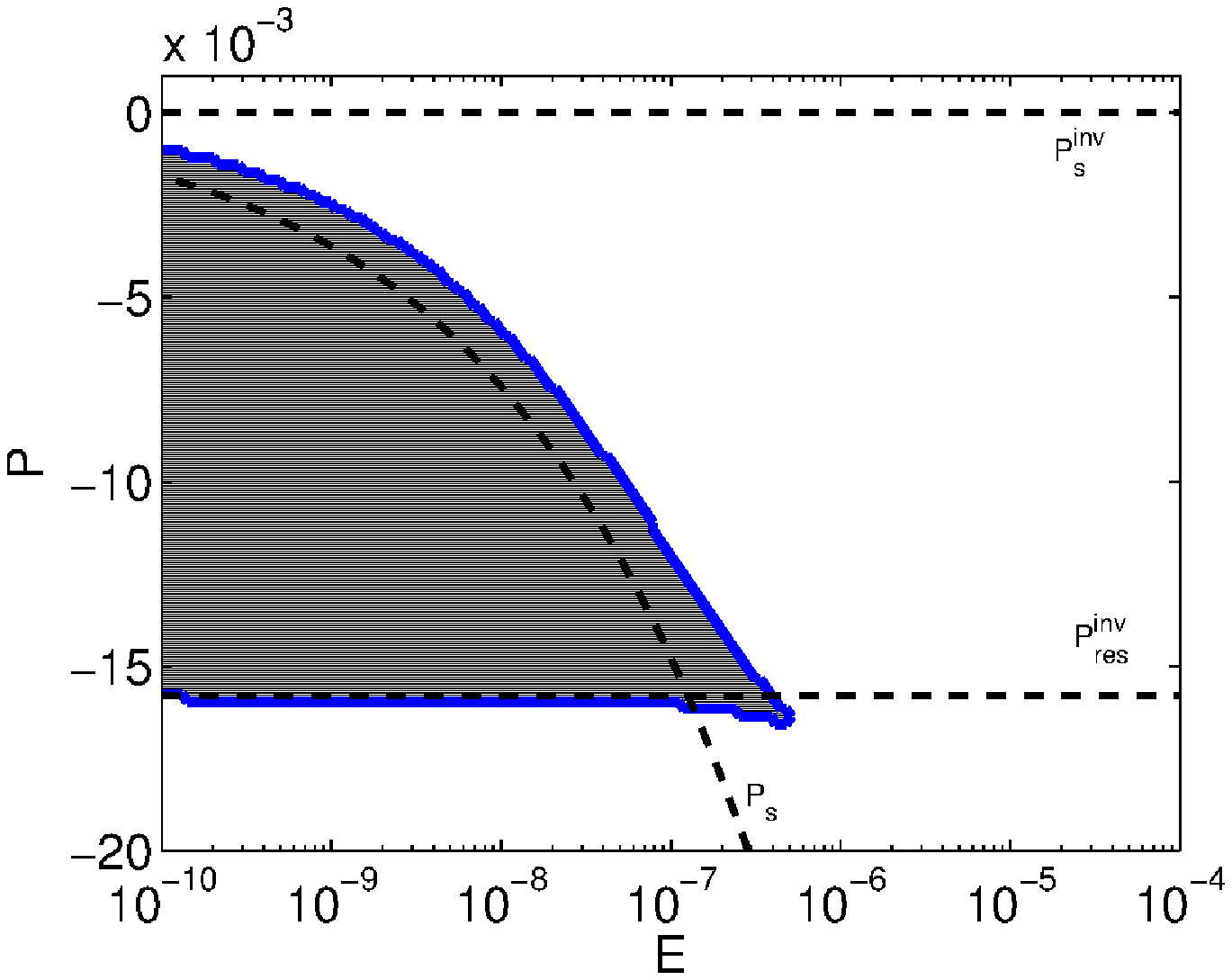}}
      \subfigure[]{\includegraphics[scale=0.47]{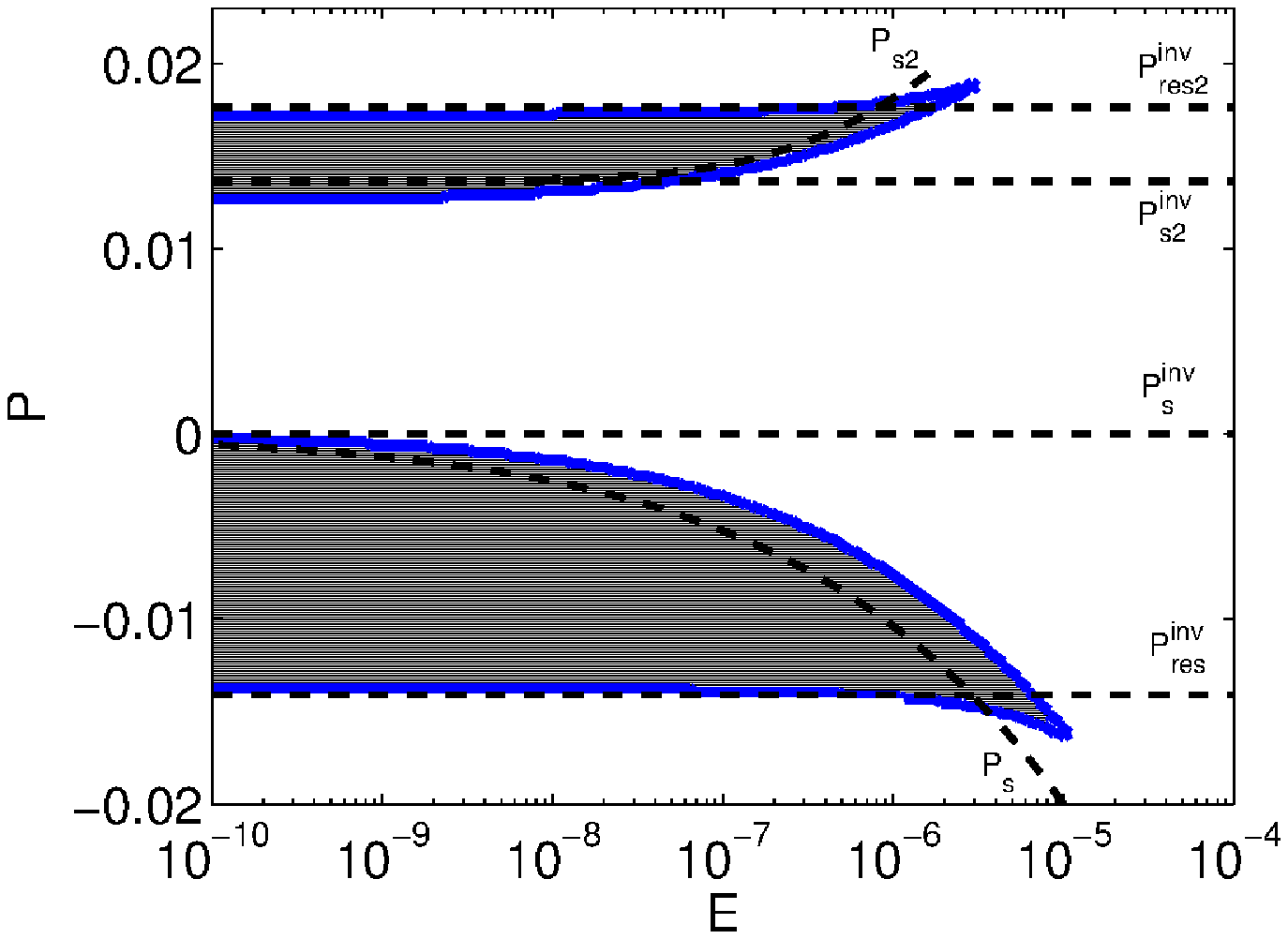}}
     \end{tabular}
\caption{Hatched areas bounded by thick solid lines represent the two multiple solutions zones of equations (\ref{eq:NoSpinUp})-(\ref{eq:Busse3}), whereas the dashed lines are our analytical estimates. Parameters: (a) small precession angle of $\alpha=2.86^{\circ}$, with $c/a=0.98$, which gives $\lambda_r \approx -2.637$ and $\lambda_i \approx 0.290$ (inviscid values obtained from the formula of Zhang and co-workers \cite{zhang2004inertial}); (b) large precession angle ($\alpha=83.7^{\circ}$), with $c/a=0.97$, which gives $\lambda_r \approx -2.646$ and $\lambda_i \approx 0.306$ (inviscid values obtained from the formula of Zhang and co-workers \cite{zhang2004inertial}). }
    \label{fig:CompaBusse}             
  \end{center}
\end{figure}

These various estimates have been represented in figure (\ref{fig:alphEmax}), showing that they capture correctly the multiple solutions zones features in the inviscid limit. Now, we can wonder how these estimates compare with the exact solutions of equations  (\ref{eq:NoSpinUp})-(\ref{eq:Busse3}). In figure \ref{fig:CompaBusse}, equations (\ref{eq:NoSpinUp})-(\ref{eq:Busse3}) are solved for two different precession angles and spheroids, and the ranges of parameters where we have multiple solutions are located between the thick solid lines. As expected, for a small precession angle (figure \ref{fig:CompaBusse}a), there is a unique multiple solutions zone, whereas two multiple solutions zones exist for a large precession angle (figure \ref{fig:CompaBusse}b, where we have added to figure \ref{fig:mult2}a the various analytical estimates we have obtained). Figure \ref{fig:CompaBusse} confirms that the analytical estimates obtained from the reduced model capture quite well these multiple solutions zones. The previously derived expressions allow thus to bound quite accurately these zones, especially in the inviscid limit.

\section{Solutions stability} \label{sec:stab}

\subsection{Estimates of the Jacobian eigenvalues}
Having localized the ranges of parameters where multiple solutions exist for the flow in a precessing spheroid, one can wonder if it is possible to observe experimentally these multiple solutions, i.e. what is the stability of these solutions. According to Noir and co-workers \cite{Noir2003}, the branch between $P_{s}$ and $P_{res}$ is unstable and cannot thus be physically realized. However, very few details are provided, and we propose thus to reinvestigate this issue here, using the dynamical model described in section \ref{sec:dynmod}.
 
  \begin{figure}
  \begin{center}
    \begin{tabular}{ccc}
      \subfigure[]{\includegraphics[scale=0.47]{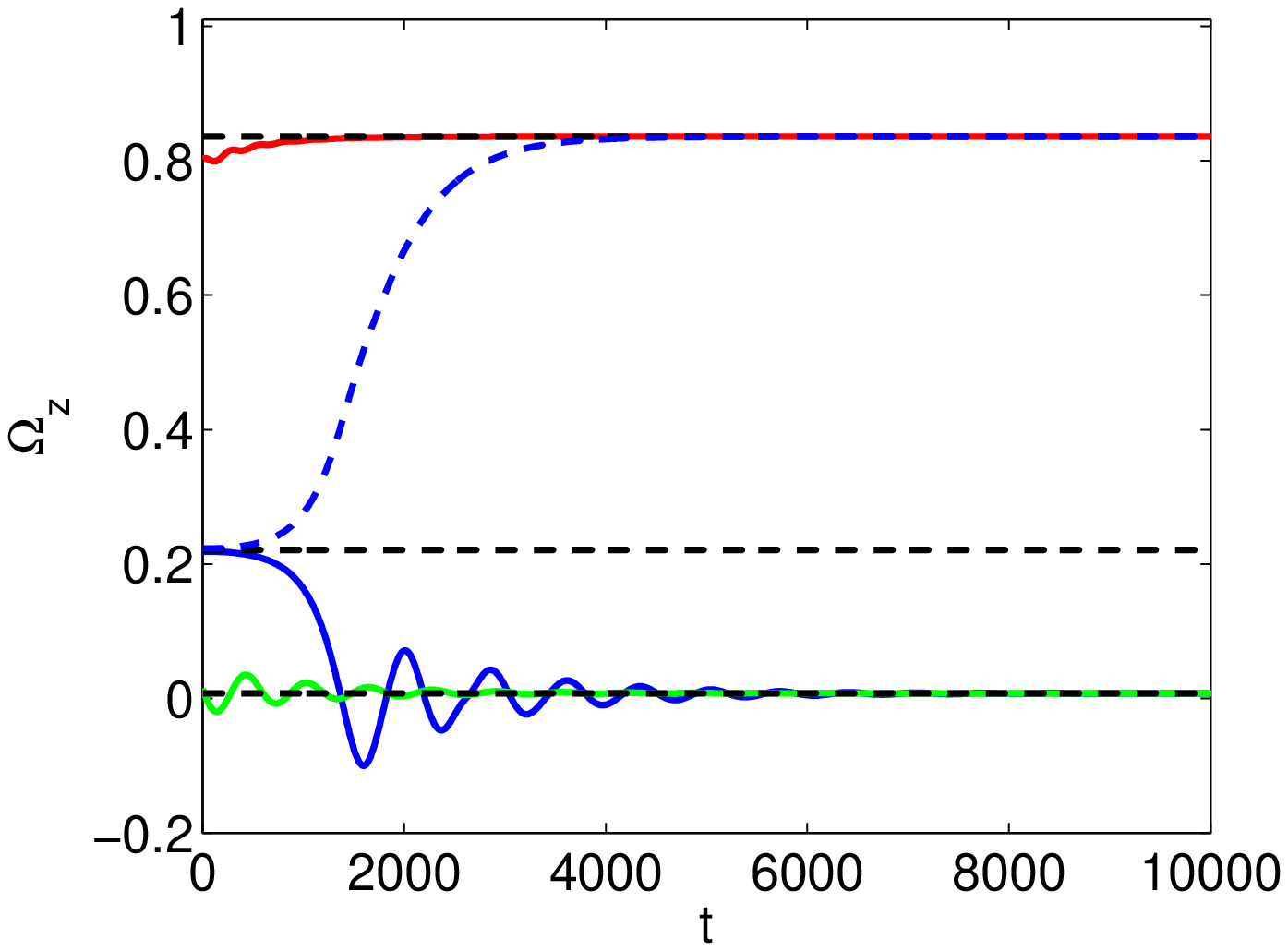}}
            \subfigure[]{\includegraphics[scale=0.47]{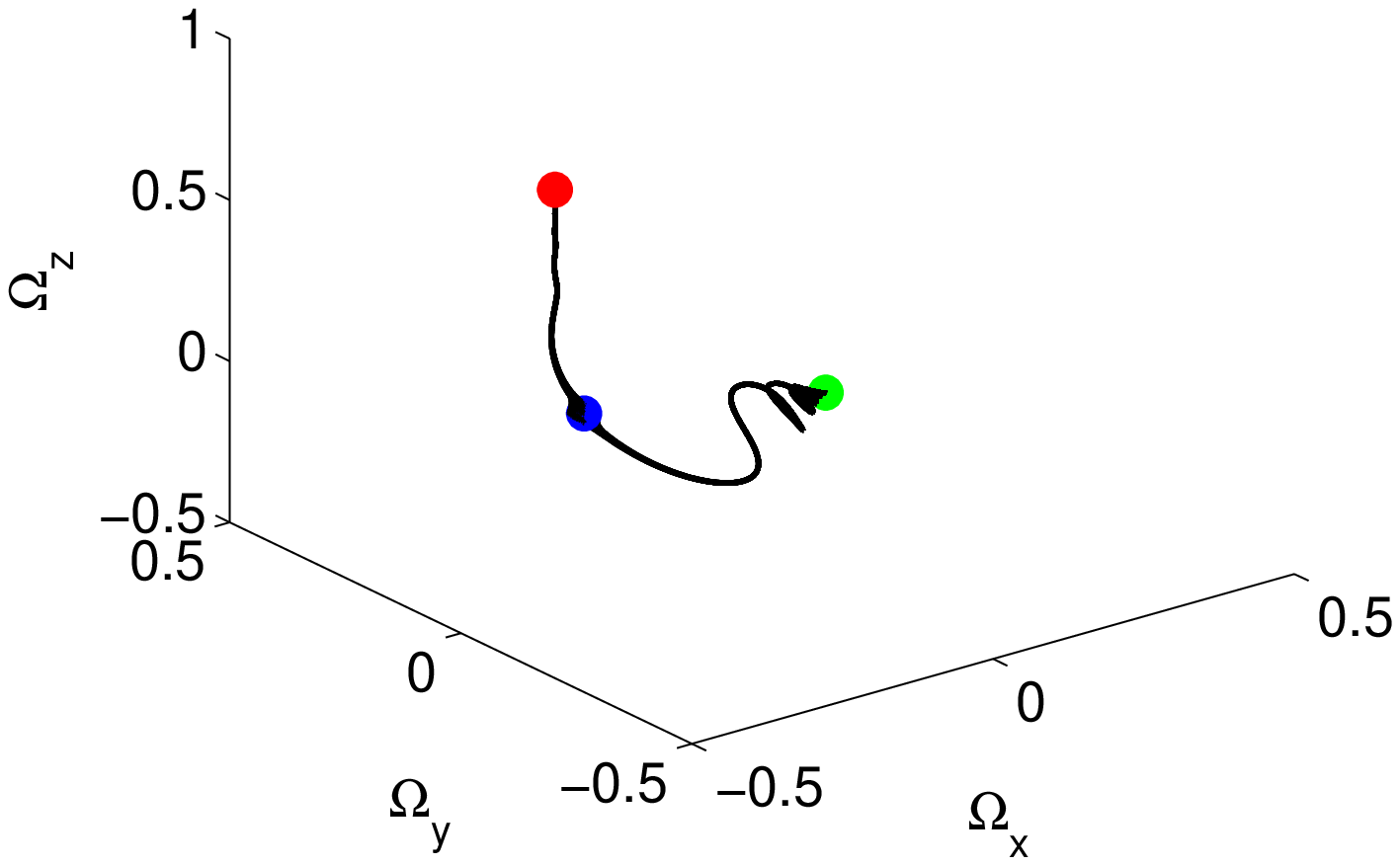}}  \\
      
      \subfigure[]{\includegraphics[scale=0.47]{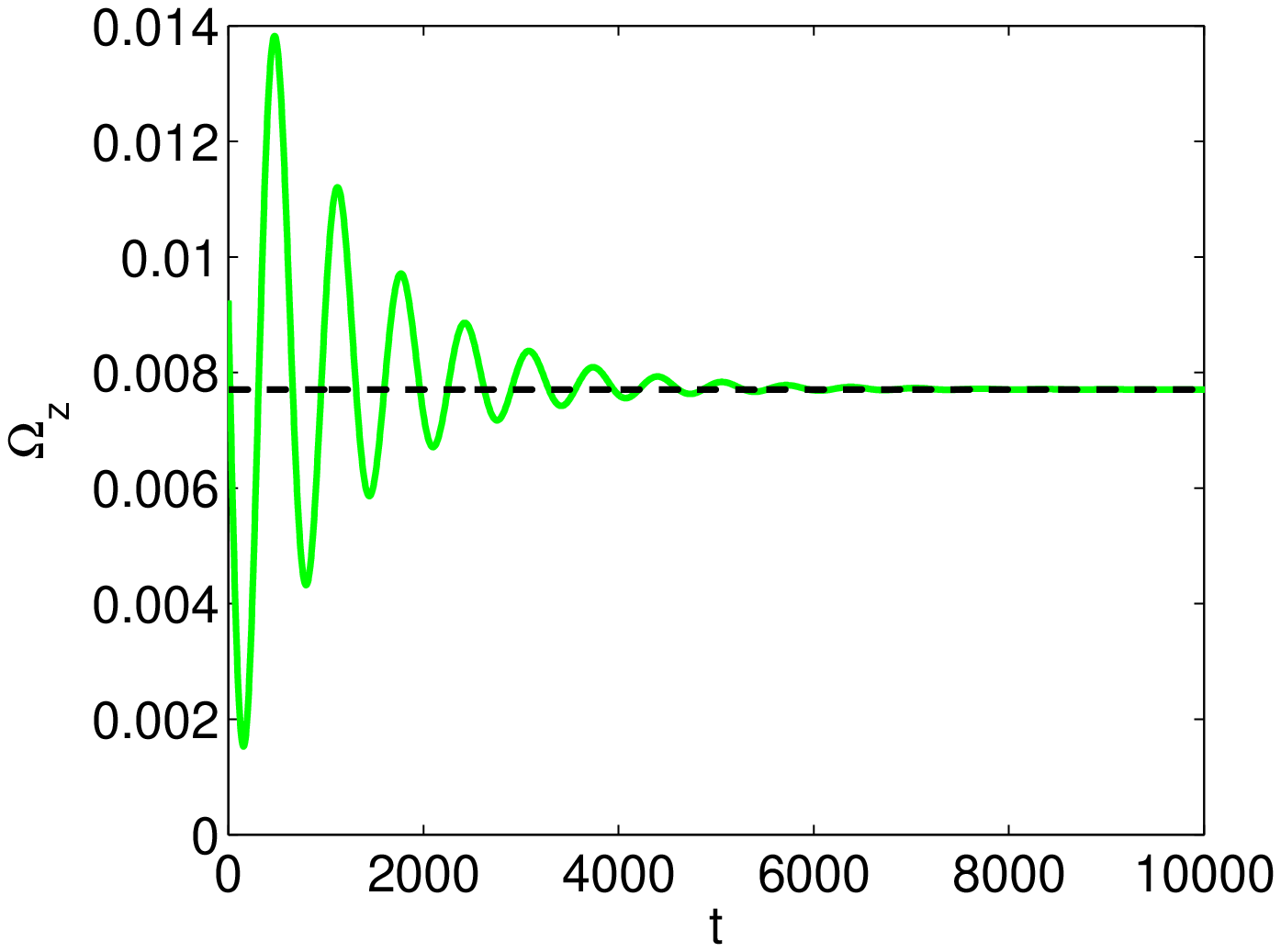}} 
            \subfigure[]{\includegraphics[scale=0.47]{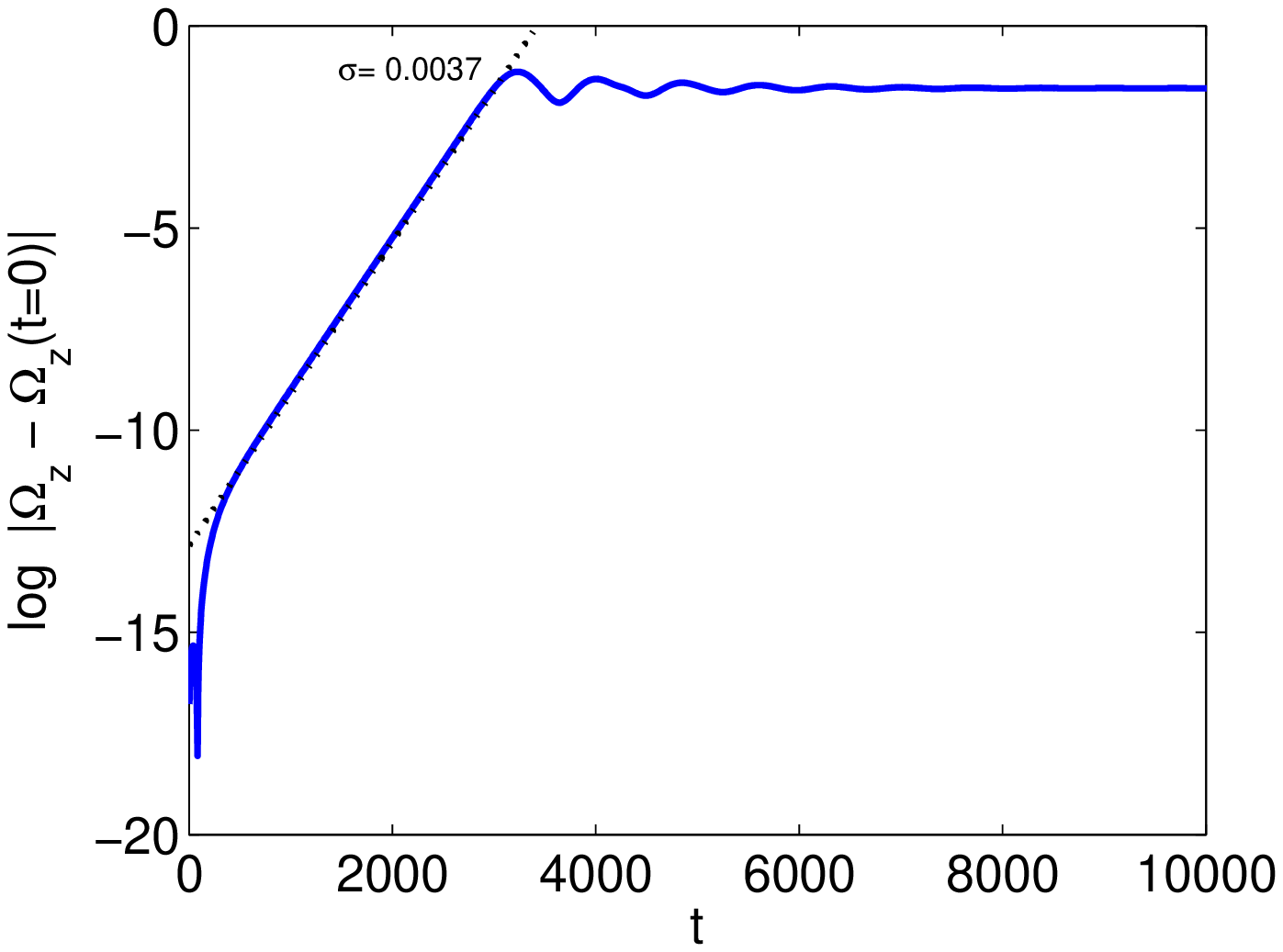}}

     \end{tabular}
\caption{(a) Time-evolution of $\Omega_z$ (solid lines), starting from the three perturbed fixed points(dashed lines) of the dynamical model (\ref{eq_axi_P_x})-(\ref{eq_axi_P_z}), with the viscous term (\ref{eqNoSUP}), relevant for equations (\ref{eq:NoSpinUp})-(\ref{eq:Busse3}), for $\alpha=81^{\circ}$, $P=-0.011$, $E=10^{-6}$, and $c/a=0.97$, which gives $\lambda_r \approx -2.646$ and $\lambda_i \approx 0.306$ (inviscid values obtained from the formula of Zhang and co-workers \cite{zhang2004inertial}). The (unstable) intermediate solution can evolve toward the other fixed point with a different initial perturbation (dashed line).  (b) Perturbing the unstable fixed point in all the directions of the space $(\Omega_x,\Omega_y,\Omega_z)$, we show here that the time-evolution of this solution systematically follows the same path in this space (circles correspond to the three fixed points). (c) Same as figure a, with only the solution starting from the lowest $\Omega(t=0)$ (stable fixed point). (d) Same as figure a, with only the solution starting from the  perturbed unstable equilibrium point. Time-evolution of $\Omega_z$ is plotted in this way to show the clear exponential growth, with a measured growth rate of $\sigma \approx 0.0037$.}
    \label{fig:CompaStab}             
  \end{center}
\end{figure}

We consider the linear stability of the equilibrium solution $\boldsymbol{\Omega}^0$, and we investigate the fate of the flow $\boldsymbol{\Omega}^0 +\boldsymbol{\epsilon}$ (where $|\boldsymbol{\epsilon}|=|(\epsilon_x,\epsilon_y,\epsilon_z)| \ll 1$). Inserting this ansatz in the dynamical reduced model (equations \ref{eq_axi_P_x}-\ref{eq_axi_P_x} and \ref{addhocvisc}) leads to 
\begin{eqnarray}
\displaystyle
\frac{\partial \boldsymbol{\epsilon}}{\partial t} =
 \left [
   \begin{array}{ccc}
      - \lambda_r \sqrt{E} & P_z (1-e) -e \Omega_z^0 & -e \Omega_y^0 \\
      -P_z (1-e) + e \Omega_z^0& - \lambda_r \sqrt{E} & P_x+e \Omega_x ^0\\
      0& -P_x (1+e) & - \lambda_r \sqrt{E} \\
   \end{array}
   \right ]\, \boldsymbol{\epsilon} =\boldsymbol{M_{r}} \boldsymbol{\epsilon},  \label{eq:Mred}
 \end{eqnarray}
 where the eigenvalues $\mu_i$ (with $i=1,\, 2,\, 3$) of the Jacobian matrix $\boldsymbol{M_{r}}$ characterize the stability of the equilibrium $\boldsymbol{\Omega}^0$. Naturally, we can obtain similar Jacobian matrix $\boldsymbol{M_{g}}$ and $\boldsymbol{M_{B}}$  using respectively the viscous term (\ref{eq:GenVisc}) of the generalized model, or its simplified form  (\ref{eqNoSUP}) in the validity limit of equations (\ref{eq:NoSpinUp})-(\ref{eq:Busse3}).

Considering  the dynamical model (\ref{eq_axi_P_x})-(\ref{eq_axi_P_z}), with the viscous term (\ref{eqNoSUP}), relevant for equations (\ref{eq:NoSpinUp})-(\ref{eq:Busse3}), we show in figure \ref{fig:CompaStab}a the time evolution of $\Omega_z$, when we start from the three perturbed fixed points (dashed lines). The solution with the intermediate initial condition $\Omega_z(t=0)$ is clearly unstable, and evolves toward one of the two other fixed points, depending on the initial perturbation. As a complementary view, we show in figure \ref{fig:CompaStab}b the time-evolution of this unstable solution in the space $(\Omega_x,\Omega_y,\Omega_z)$, when exploring all the possible perturbations. Note that the  time-evolution of this solution systematically follows the same path in this space.

Figure \ref{fig:CompaStab}c is a zoom of figure \ref{fig:CompaStab}a, showing only the solution starting from the smallest  $\Omega_z(t=0)$. This solution comes back to the fixed point as a damped harmonic oscillator, decaying exponentially towards the equilibrium point at a decay rate of $\sigma \approx -8.9 \cdot 10^{-4}$, and oscillating at the frequency $9.6 \cdot 10^{-3}$. For this solution, the three eigenvalues $\mu_i$ of $\boldsymbol{M_{B}}$ are $\mu_1\approx -0.63 \cdot 10^{-3}$, $\mu_2=(-0.88+9.5 \textrm{i}) \cdot 10^{-3}$, and $\mu_3 =\overline{\mu_2}$, complex conjugate of $\mu_2$. The real parts of the three eigenvalues are all negative, confirming the stability of the fixed point, and the eigenvalues are in $1 \%$ agreement with the measured  decay rate and oscillation frequency. Using the eigenvectors, we have naturally chosen here the right initial perturbation to test $\mu_2$, but the three eigenvalues can actually be checked with the differential equations solution, by calculating numerically $\boldsymbol{M_{r}}$ at $t=0$, using $\partial \boldsymbol{\epsilon}/\partial t$. As expected, this approach gives very close eigenvalues.

  \begin{figure}
  \begin{center}
    \begin{tabular}{ccc}
      \subfigure[]{\includegraphics[scale=0.47]{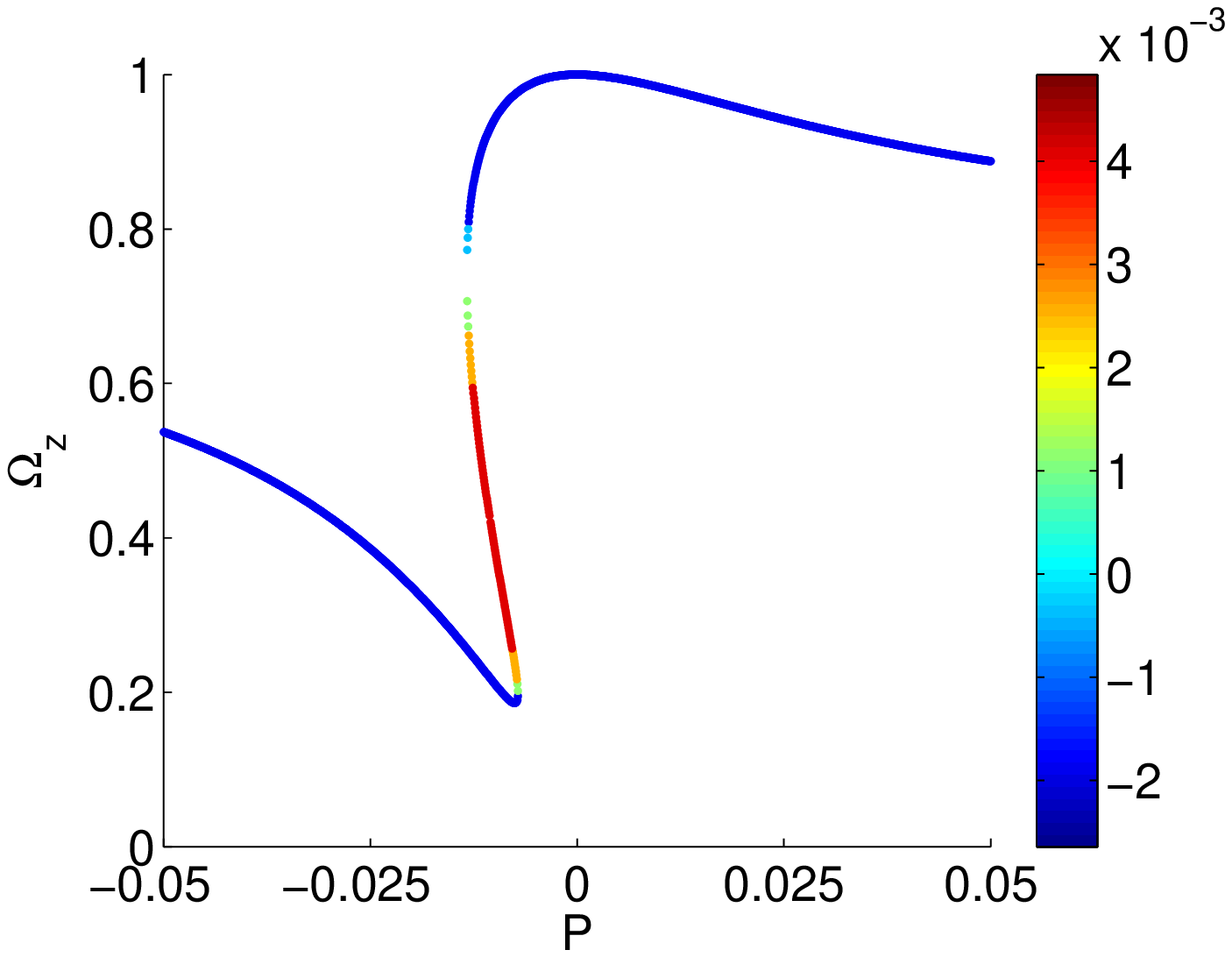}}
      \subfigure[]{\includegraphics[scale=0.47]{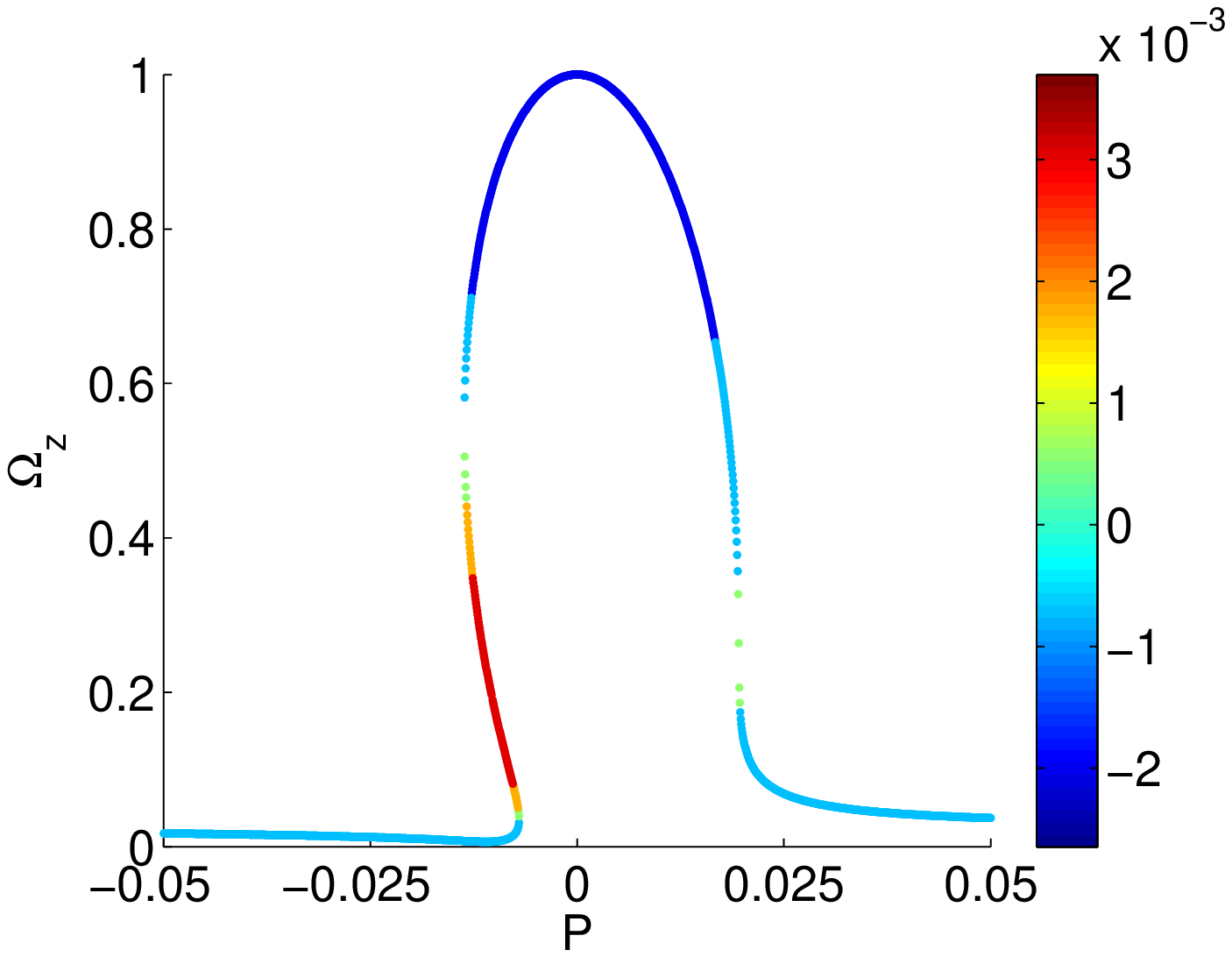}}
     \end{tabular}
\caption{Fixed points of (\ref{eq_axi_P_x})-(\ref{eq_axi_P_z}), with the viscous term (\ref{eqNoSUP}), for $\alpha=81^{\circ}$, $E=10^{-6}$, and $c/a=0.97$, which gives $\lambda_r \approx -2.646$ and $\lambda_i \approx 0.306$ (inviscid values obtained from the formula of Zhang and co-workers \cite{zhang2004inertial}). The colorbar shows $\max_i \mathcal{R}e(\mu_i)$, i.e. the maximum of the real part of the eigenvalues, indicating unstable solutions when positive. (a) $\alpha=30^{\circ}$.  (b) $\alpha=81^{\circ}$. It is clear than the fixed points are unstable only on the branch linking $P_s$ to $P_{res}$, and on the branch linking $P_{s2}$ to $P_{res2}$.}
    \label{fig:CompaStab2}             
  \end{center}
\end{figure}
Figure \ref{fig:CompaStab}d is a zoom of figure \ref{fig:CompaStab}a, showing only the solution starting from the unstable fixed point. The solution shows a clear exponential departure from the equilibrium point, with a growth rate $\sigma \approx  3.7 \cdot 10^{-3}$, and an oscillation frequency of $7.8 \cdot 10^{-3}$. For this solution, the three eigenvalues of $\boldsymbol{M_{B}}$ are $\mu_1\approx 3.73 \cdot 10^{-3}$, $\mu_2=(-4.6+5.7 \textrm{i}) \cdot 10^{-3}$, and $\mu_3 =\overline{\mu_2}$. Since the real part of one eigenvalue is positive, this fixed point is confirmed to be unstable, and the growth rate is in excellent agreement with the theoretical prediction (oscillation frequencies are in less good agreement).

Finally, the eigenvalues for the last solutions are $\mu_1 \approx -2 \cdot 10^{-3}$,  $\mu_2=(-0.28+2.4 \textrm{i}) \cdot 10^{-2}$, and $\mu_3 =\overline{\mu_2}$, in excellent agreement with the eigenvalues obtained from the time-evolution of the solutions.

In figure \ref{fig:CompaStab2}, we consider the same parameters as figure \ref{fig:CompaStab}, and we vary the Poincar\'e number $P$, plotting $\max_i \mathcal{R}e(\mu_i)$ as a colorbar to indicate the stability of each solution. It allows to clearly see that the fixed points are unstable only on the branch linking $P_s$ to $P_{res}$, and on the branch linking $P_{s2}$ to $P_{res2}$. In the particular case $\alpha < \alpha_{lim}$, we thus recover the conclusion of Noir and co-workers \cite{Noir2003}.

One can also investigate analytically, in particular cases, the eigenvalues of the Jacobian matrix. Considering for instance the matrix (\ref{eq:Mred}), it is clear that this requires an analytical estimate of $\Omega_z^0$ (with $\Omega_x^0$, $\Omega_y^0$ respectively given by equations \ref{eq:Wx}, \ref{eq:Wy}).  The full analytical solutions (given by the roots of equation \ref{eq:appEQ}) are very lengthy and are thus not tractable, and we will thus work in two simplifying limit cases. The first limit case is  $\alpha \ll 1$, where
\begin{eqnarray}
 \mkern-72mu \Omega_z=1+\alpha^2 \left[ \frac{(1+e)(K^2-e^2)}{e^4}P^2+2 \frac{(1-e^2)(2K^2-e^2)}{e^5}P^3 \right]+\mathcal{O}(P^4+K^3)
\end{eqnarray}
with $K=-\lambda_r \sqrt{E}$. The second limit case is $\alpha=\pi/2$, where 
\begin{eqnarray}
 \mkern-72mu \Omega_z = \left|
 \begin{array}{c}
 \displaystyle
     |P| <P_{res}=\frac{e}{2\sqrt{1+e}} :  \, \, \, \,   \, \,  \Omega_z^l=\frac{K^2}{(1+e) P^2} \\
       \\
        \displaystyle
     |P| <P_{res} \, \,  \& \, \,  E<E_{max} :  \, \, \, \,   \, \, \Omega_z^*=\frac{(1+e) P^2}{e^2} \left(1-\frac{K^2}{e^2} \right)-\frac{K^2}{(1+e)P^2} \\
      \\
       \displaystyle
   |P| <P_{res}: \, \, \, \,   \, \,   \Omega_z^u=1- \left[ \frac{1+e}{e^2}-\frac{1+e}{e^4} K^2 \right] P^2 \\
   \end{array}
\right. , \label{eq:Wz90} 
\end{eqnarray}
at the order $\mathcal{O}(P^4+K^3)$. In this case, we have clearly three possible solutions, the lower one $\Omega_z^l$, the unstable one $\Omega_z^* $, and the upper one $\Omega_z^u$.

  \begin{figure}
  \begin{center}
    \begin{tabular}{ccc}
      \subfigure[]{\includegraphics[scale=0.47]{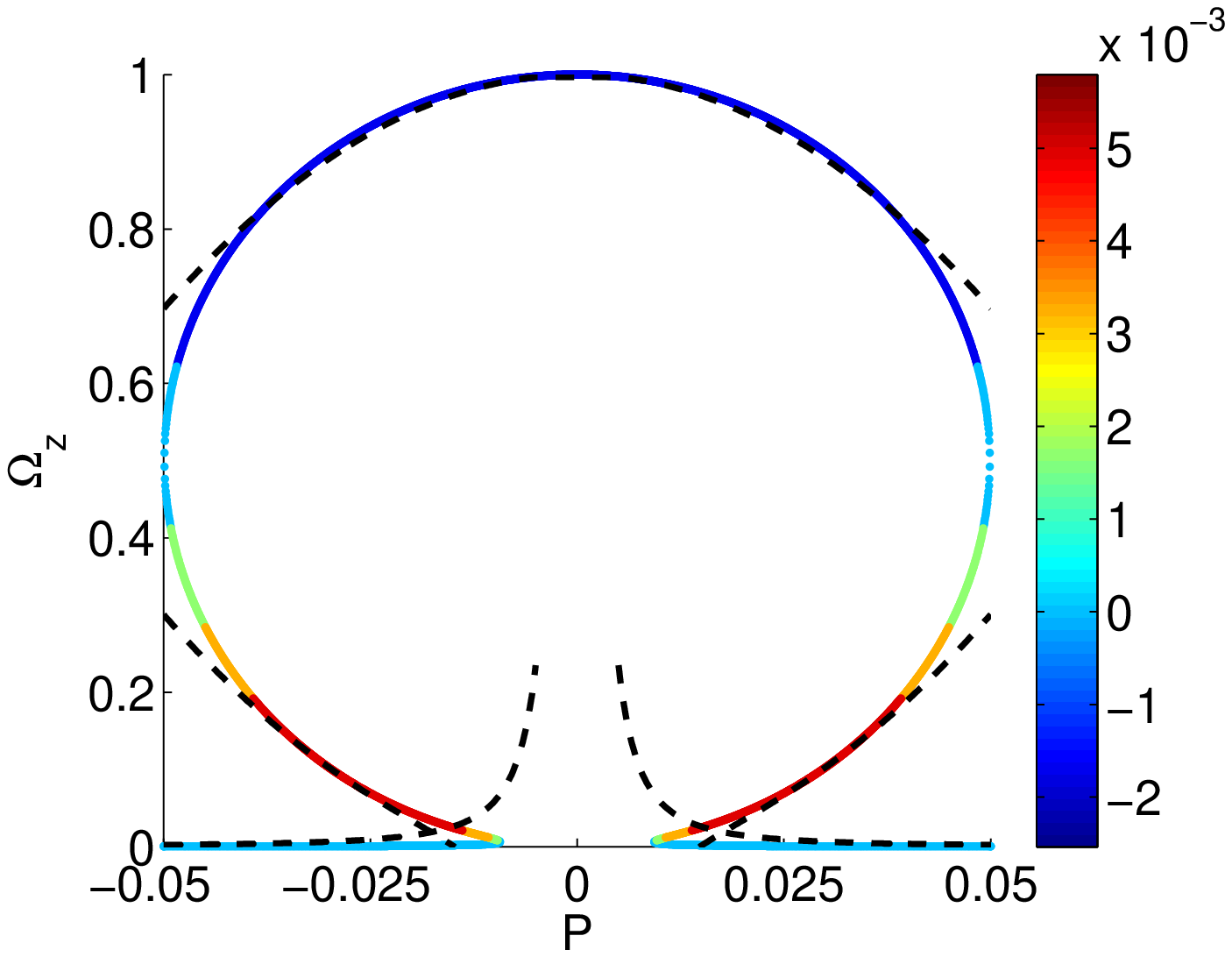}}
      \subfigure[]{\includegraphics[scale=0.47]{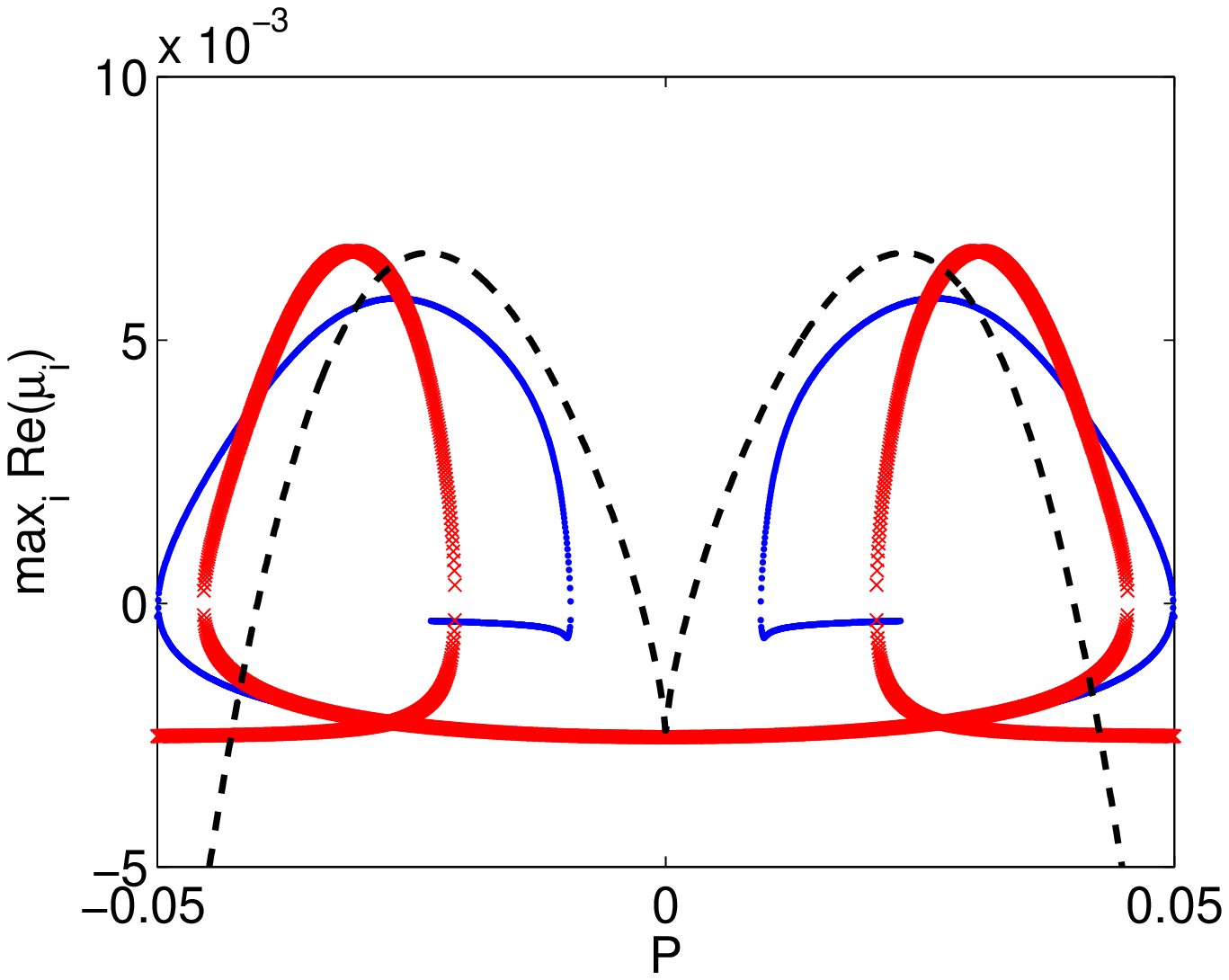}}
     \end{tabular}
\caption{Fixed points of (\ref{eq_axi_P_x})-(\ref{eq_axi_P_z}) for $\alpha=90^{\circ}$, $E=10^{-6}$, and $c/a=1.1$, which gives $\lambda_r \approx -2.536$ and $\lambda_i \approx 0.120$ (inviscid values obtained from the formula of Zhang and co-workers \cite{zhang2004inertial}). (a) With the viscous term (\ref{eqNoSUP}).  The colorbar shows $\max_i \mathcal{R}e(\mu_i)$, and the dashed lines are given by equation (\ref{eq:Wz90}), obtained with the reduced viscous term (\ref{addhocvisc}). (b) Dots are calculated with the viscous term (\ref{eqNoSUP}), crosses with the reduced viscous torque  (\ref{addhocvisc}), and the dashed line corresponds to equation (\ref{eq:mus}).}
    \label{fig:muU}             
  \end{center}
\end{figure}

Based on these expansion, one can estimate the eigenvalues in the limit $\alpha \ll 1$, 
\begin{eqnarray}
\mu_1 &=& -K , \\
\mu_2 &=& -K+\textrm{i}(1-e) (P-P_r), \\
\mu_3&=& \overline{\mu_2}
\end{eqnarray}
where $P_r$ is the (linear) Poincar\'e resonance (see equation \ref{eq:Pres}). Here, the solution is stable, and, if perturbed, the oscillation frequency $|\mathcal{I}m\, \mu_2|=|\mathcal{I}m\, \mu_3|$ is proportional to $P-P_r$. Note that the third eigenvalue is systematically the complex conjugate of $\mu_2$, and is thus not considered hereinafter. 

  In the limit $\alpha=\pi/2$, we have 
\begin{eqnarray}
\mu_1^l &=& -K  , \\
\mu_2^l&=& -K+\textrm{i}P \sqrt{1+e},
\end{eqnarray}
for $\Omega_z^l$, and
\begin{eqnarray}
\mu_1^* &=& -K +[(1+e) K P^2]^{1/3}-\frac{2}{3e^2}[(1+e)^5 P^{10}/K]^{1/3} , \label{eq:mus} \\
\mu_2^* &=& -K-\frac{[(1+e) K P^2]^{1/3}(1-\textrm{i}\sqrt{3})}{2} +\frac{[(1+e)^5 P^{10}/K]^{1/3} (1+\textrm{i}\sqrt{3})}{3e^3} ,
\end{eqnarray}
for $\Omega_z^*$
\begin{eqnarray}
\mu_1^u&=& -K +(1+e)KP^2/e^2 , \\
\mu_2^u&=& -K+\textrm{i}[e-P^2(1+1/e)],
\end{eqnarray}
for $\Omega_z^u$. The solution $\Omega_z^*$ becomes unstable due to the driving term $\chi=[(1+e) K P^2]^{1/3}$ in equation (\ref{eq:mus}), which is thus the control parameter of this instability. Note that the real part of the two other eigenvalues of $\Omega_z^*$ is always negative, denoting a saddle-node bifurcation of $\Omega_z^*$.

In figure \ref{fig:muU}a, we consider the limit case $\alpha=90^{\circ}$, which is a perfectly symmetrical configuration for $P$ and $-P$. In figure \ref{fig:muU}, the growth rate/decay of the dynamical model corresponding to the Busse equation are shown, and compared with the results of the reduced model. The estimate (\ref{eq:mus}) is also shown, capturing correctly the global behaviour of the growth rate/decay, with a maximum reached for a Poincar\'e number of
\begin{eqnarray}
P_{max}=\pm \frac{1}{\sqrt{(1+e) K}} \left( \frac{3e^2 K^2}{10} \right)^{3/8},
\end{eqnarray}
giving an analytical estimate of the most unstable Poincar\'e number. 

One can wonder if we can interpret the previous results in the framework of the dynamical systems theory, and, for instance, obtain typical bifurcation diagrams. In section \ref{eq:bifurc}, we focus on the case $\alpha=\pi/2$, where exact calculations can be performed analytically without any expansion or hypothesis, in order to investigate the onset, i.e. the zone around $E_{max}$.

 \subsection{Nature of the bifurcation} \label{eq:bifurc}
Considering the cubic equation governing the fixed points of $\Omega_z$, given by equation (\ref{eq:appEQ}), one can obtain easily obtain $P=f(e,K,\Omega_z)$ as roots of a quadratic equation (without any hypothesis). Equation $\partial P/ \partial \Omega_z=0$, which defines $P_{res}=\partial \Omega_z/ \partial P = \infty$, is, in the general case, a polynomial of degree $6$. However, it reduces to a cubic polynomial for $\alpha=\pi/2$, given by
\begin{eqnarray} \label{eq:cubic90}
-2 e^2 \Omega_z^3+e^2 \Omega_z^2-K^2=0.
\end{eqnarray}
Note that, for $K=0$, we obtain  $\Omega_z^{res}=1/2$ and $\Omega_z^{res}=0$, which naturally allows to recover $P=f(e,K,\Omega_z)=e/(2\sqrt{1+e})=P_{res}$. The discriminant $\Delta^{res}$ of the polynomial (\ref{eq:cubic90}) is such that
\begin{eqnarray} \label{eq:discr90}
\Delta^{res}=\frac{K^2(27 K^2-e^2)}{108 e^4} =0 \Longleftrightarrow K=-\lambda_r E=\frac{e}{3^{3/2}}.
\end{eqnarray}
Then, when $\lambda_r^2 E\geq e^2/27$, equation (\ref{eq:cubic90}) have only one real solution, and, at most, three solutions otherwise. Since we are considering the resonant Poincar\'e number $P$, this shows that we have obtained the rigorous value $\lambda_r^2 E_{max}=e^2/27$  for $\alpha=\pi/2$. At this particular point  $ K=e/(3^{3/2})$, the three roots of equation (\ref{eq:cubic90}) are equal to $\Omega_z^{E_{max}}=1/3$, and $P$ is given by 
\begin{eqnarray}
P_{res}^{E_{max}}=f(e,K,\Omega_z)=\frac{1}{2} \sqrt{\frac{32}{27}}\, \frac{e}{\sqrt{1+e}}.
\end{eqnarray}

\begin{figure}
  \begin{center}
    \begin{tabular}{ccc}
      \subfigure[]{\includegraphics[scale=0.47]{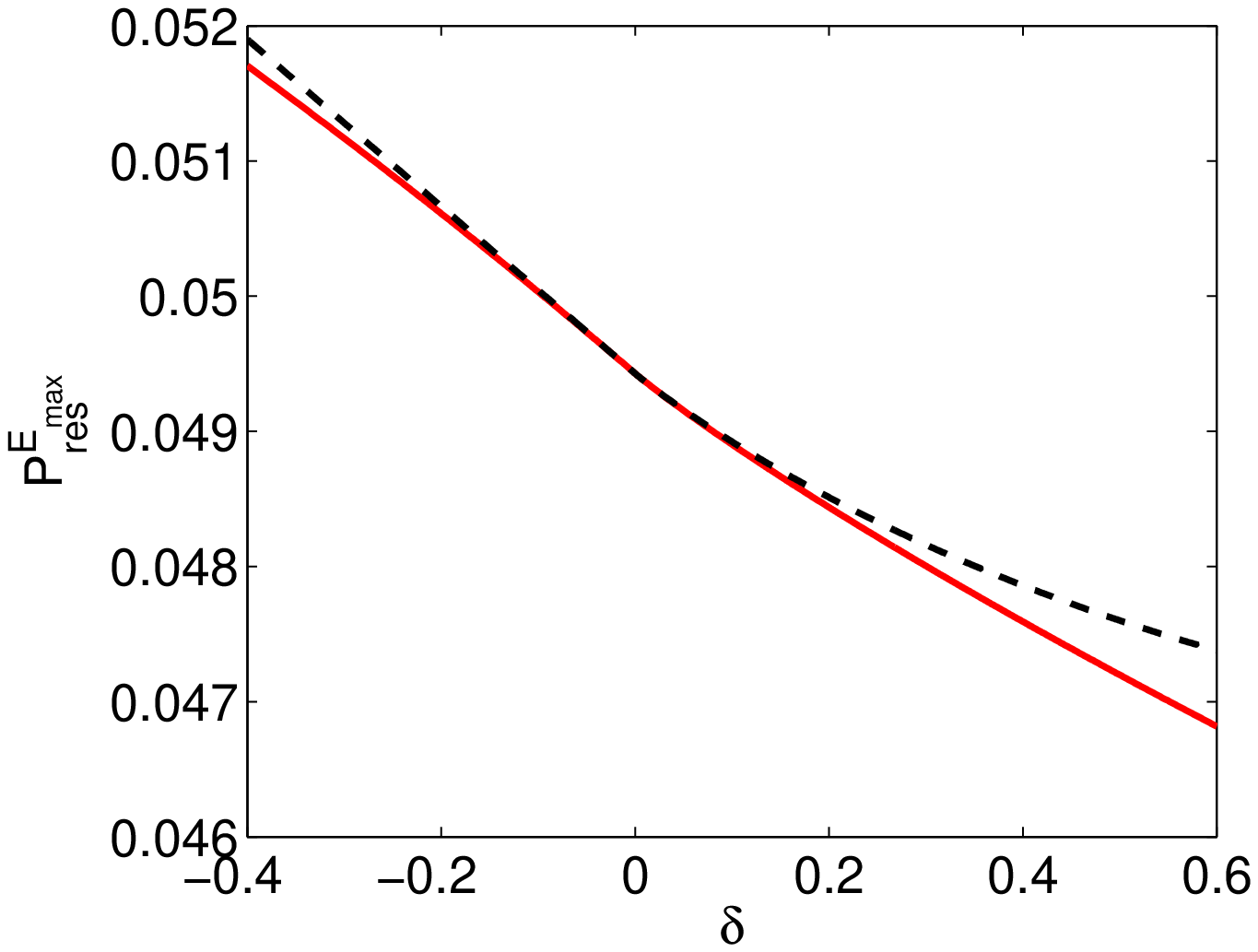}}
      \subfigure[]{\includegraphics[scale=0.47]{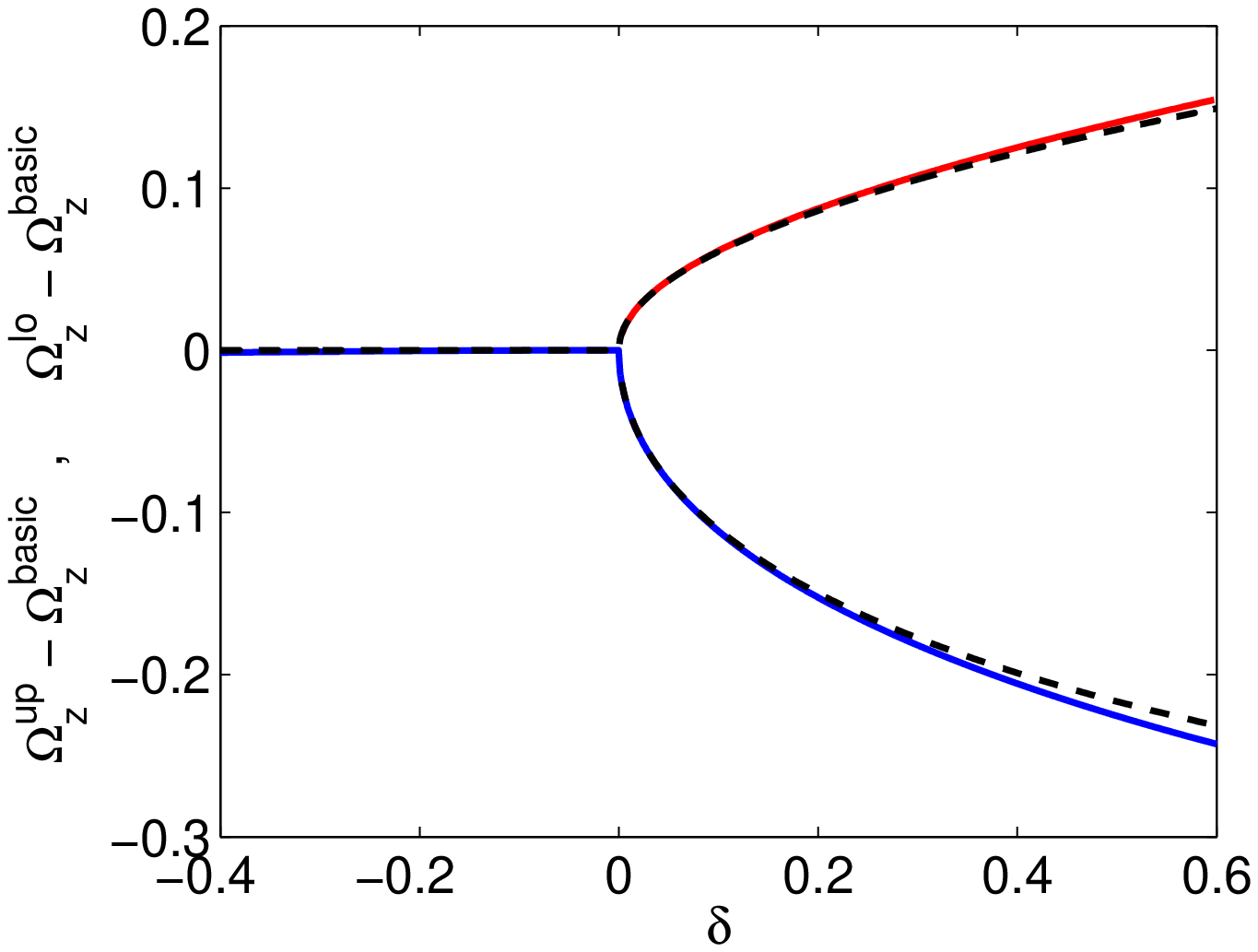}}
     \end{tabular}
\caption{Results for the fixed points of (\ref{eq_axi_P_x})-(\ref{eq_axi_P_z}) with the reduced viscous term (\ref{addhocvisc}), using $c/a=1.1$, $\alpha=\pi/2$, and the perturbed Ekman number (\ref{eq:Emax90e}), such that $K=e (1-\delta/2)/(3^{3/2})$. (a) Comparison of the exact $P_{res}$ (solid line) with the expansion (\ref{eq:Pres90e}). (b) Comparison of the two exact solutions, $\Omega_z^{up} $ and $\Omega_z^{lo} $ (solid lines), bifurcating from the unique solution $\Omega_z^{basic}$ (existing for $\delta \leq 0$)  with their respective expansions (\ref{eq:Bif1}), and (\ref{eq:Bif2}).}
    \label{fig:pitch}             
  \end{center}
\end{figure}

We now focus on the onset of the instability by perturbing $ E_{max}$ into the perturbed Ekman number
\begin{eqnarray}
E_{max}=\frac{e^2}{27 \lambda_r^2}\, (1-\delta), \label{eq:Emax90e}
\end{eqnarray}
which is equivalent to consider the perturbed quantity $K=e (1-\delta/2)/(3^{3/2})$. We then have
\begin{eqnarray}
P_{res}^{E_{max}}=f(e,K,\Omega_z)=  \frac{e}{36 \sqrt{1+e}}\left[ \sqrt{384} -\sqrt{6}\, \delta + \sqrt{2} \, \delta^{3/2} \right] +\mathcal{O}(\delta^{2}) . \label{eq:Pres90e}
\end{eqnarray}
When $\delta\leq0$, the unique solution of equation (\ref{eq:cubic90}) is thus
\begin{eqnarray}
\Omega_z^{basic} = \frac{1}{3}-\frac{1}{27}\, \delta ,
\end{eqnarray}
at leading order in $\delta$. This corresponds to the basic state. Note that, for $\delta\leq0$, we naturally do not have $\partial \Omega_z/ \partial P = \infty$ anymore ; this basic state simply corresponds to the flow obtained when $P$ is given by equation (\ref{eq:Pres90e}) and $E$ by equation (\ref{eq:Emax90e}). When $\delta\geq0$,  equation (\ref{eq:cubic90}) has the two following solutions ,
\begin{eqnarray}
\Omega_z^{up} &=& \Omega_z^{basic}+ \frac{\sqrt{3}}{9}\, \sqrt{\delta} \,  \label{eq:Bif1}  \\
\Omega_z^{lo} &=& \Omega_z^{basic}- \frac{2\sqrt{3}}{9}\, \sqrt{\delta}+\frac{1}{9}\, \delta , \label{eq:Bif2}
\end{eqnarray}
at the order $\mathcal{O}(\delta^{5/4})$. Note that the two solutions, the upper one $\Omega_z^{up}$ and the lower one $\Omega_z^{lo}  $, are not symmetrical with respect to the basic state $\Omega_z^{basic}$. In figure \ref{fig:pitch}, we compare the expansions (\ref{eq:Pres90e}), (\ref{eq:Bif1}), and (\ref{eq:Bif2}), with their exact counterparts. The agreement is very good, and one can notice the clear bifurcation of the solution $\Omega_z=1/3$ into the two solutions $\Omega_z^{up} $ and $\Omega_z^{lo} $ for $\delta>0$.

One can now calculate how the eigenvalues vary with $\delta$ by using the Jacobian matrix (\ref{eq:Mred}). We obtain, at the order $\delta^{3/2}$,
\begin{eqnarray}
 \frac{\mu_1^{up}}{e}  &=& 0 , \label{eq:bif} \\
 \frac{\mu_2^{up}}{e} &=&  \frac{\overline{\mu_3^{up}}}{e} = -\frac{\sqrt{3}}{6}+\textrm{i}\, \left( \frac{\sqrt{69}}{18}  +  \frac{611 \sqrt{23}+1656 \sqrt{3}}{69(611+72 \sqrt{69})}\, \sqrt{\delta} \right) ,
\end{eqnarray}
for the solution $\Omega_z^{up}$,  and
\begin{eqnarray}
 \frac{\mu_1^{lo}}{e} &=& - \frac{ \sqrt{3}}{8}\, \delta , \\
 \frac{\mu_2^{lo}}{e} &=&  \frac{\overline{\mu_3^{lo}}}{e} = -\frac{\sqrt{3}}{6}+\textrm{i}\, \left( \frac{\sqrt{69}}{18}  -  2\, \frac{611 \sqrt{23}+1656 \sqrt{3}}{69(611+72 \sqrt{69})}\, \sqrt{\delta} \right)  ,
\end{eqnarray}
for the solution $\Omega_z^{lo} $. At the considered order,  the solution $\Omega_z=1/3$ bifurcates for $\delta>0$ into the marginally stable solution $\Omega_z^{up} $ and the stable solution $\Omega_z^{lo} $.

To summarize this study at $\alpha=\pi/2$, the fixed point of the dynamical system (\ref{eq_axi_P_x}-\ref{eq_axi_P_z}) with the viscous term (\ref{addhocvisc}) loses stability when $\lambda_r^2 E_{max}<e^2/27$, as the real eigenvalue (\ref{eq:bif}) crosses $0$, which indicates a pitchfork bifurcation.  Besides, one can notice the usual square root dependency of the amplitude above the onset (equations \ref{eq:Bif1} and \ref{eq:Bif2}), which is clear in figure \ref{fig:pitch}. The two other eigenvalues are complex congugates with a negative real part, indicating that the pitchfork bifurcation originates from a saddle-node.

\section{Discussion}

  \begin{figure}
  \begin{center}
    \begin{tabular}{ccc}
      \subfigure[]{\includegraphics[scale=0.47]{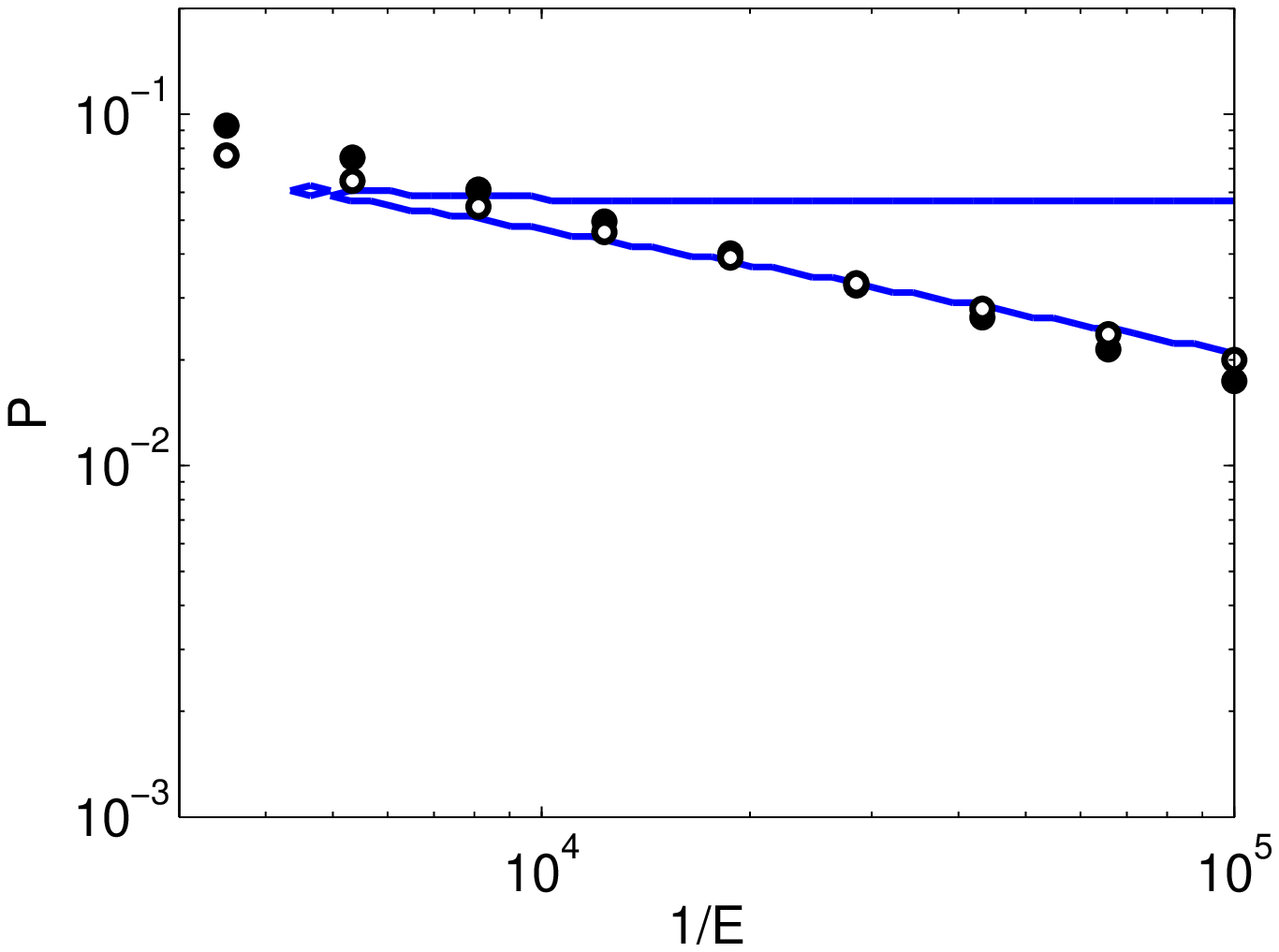}}
      \subfigure[]{\includegraphics[scale=0.47]{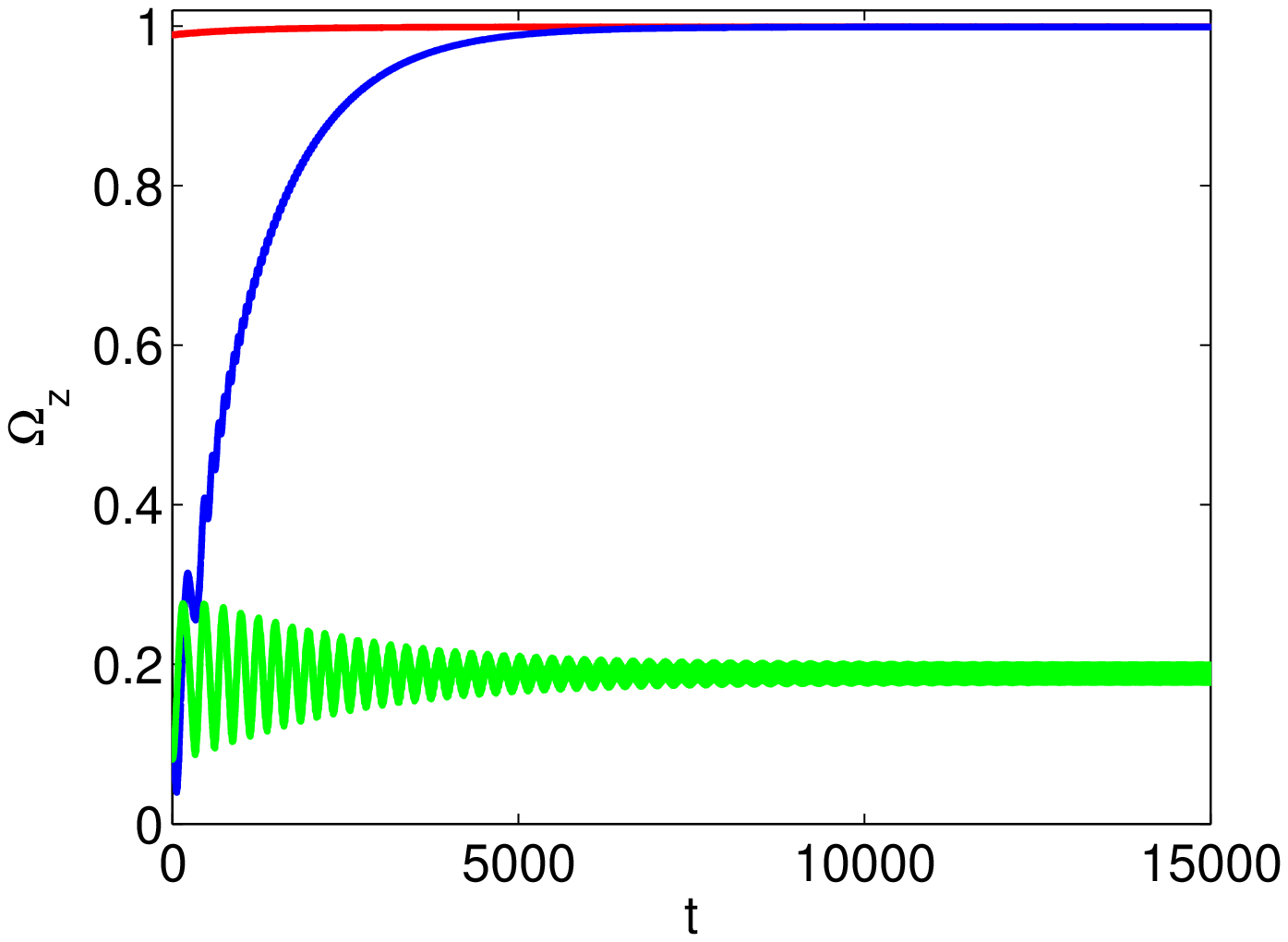}}
     \end{tabular}
\caption{(a) Considering the figure 11 of the experiments \cite{Goto2014} in a spheroid for $\alpha=\pi/2$, and $a/c=0.9$, which gives $\lambda_r \approx -2.738$ and $\lambda_i \approx 0.404$ (inviscid values obtained from \cite{zhang2004inertial}), the rounds indicates the two scaling laws, in $ P \sim E^{0.4}$ and $P \sim \sqrt{E}$,  proposed by \cite{Goto2014} for the experimental instability onset ($P=2\, E^{0.4}$ for the empty rounds, and $P=5.5\, \sqrt{E}$ for the solid rounds, determined by fitting their experimental results). The solid line represent the multiple solutions zone for equations (\ref{eq:NoSpinUp})-(\ref{eq:Busse3}), without any adjustable parameter. According to estimate (\ref{eq:Es}), the lower solid line scales as $E^{1/3}$ when $\alpha \neq \pi/2$, and as $E^{1/4}$ when $\alpha=\pi/2$ (eq. \ref{eq:NoSpinUp}-\ref{eq:Busse3}, which do not use any approximation, rather give $E^{1/3}$ here, probably because $\lambda_i \neq 0$ in these equations). (b) Considering a triaxial ellipsoid ($b/a=0.8$, $c/a=0.7$, $E=10^{-7}$, $\alpha=0.3$, $P=-0.0255$) and using the reduced viscous term (\ref{addhocvisc}), we start from the three possible steady solutions for  $\boldsymbol{\Omega}$ obtained in the spheroid with the same $c/a$. The time-evolutions clearly show that two possible flows can exist, and both are periodic in time and remain close from the stable solutions of the spheroid.}
    \label{fig:kida}             
  \end{center}
\end{figure}

We can conclude from the previous sections that two stable solutions can coexist on the branches linking $P_s$ to $P_{res}$, and $P_{s2}$ to $P_{res2}$. However, to the knowledge of the author, these coexistent solutions have never been observed in the literature, neither experimentally, nor in numerical simulations. In presence of a strong enough noise, the dynamical system could jump intermittently from one stable fixed point to the other. In this case, this could have been interpreted as an instability. In figure \ref{fig:kida}, we present the very recent experimental results obtained by Goto and co-workers \cite{Goto2014}, and we compare them to the multiple solutions zones given by equations (\ref{eq:NoSpinUp})-(\ref{eq:Busse3}). A rather good agreement is found, without any adjustable parameter. Note that this apparent good agreement could also be linked with the fact that, around $P_{res}$, the solution suddenly jumps to another branch. This drastically modifies the flow, and the new flow may then excite an instability (e.g. an inertial one, as described by \cite{kerswell1993instability}).

One can wonder if the Earth has undergone, during its evolution, parameters allowing states with possible coexistent flows. Using the values given by \cite{touma1994evolution} for a Earth-Moon distance varying between its current value and half of this value, and assuming a constant flattening equal to its current value $e\approx -0.003$, we obtain that $P$ remains of the order $P \approx -10^{-7}$, which is larger than $P_s \in[-10^{-5}; -10^{-6}]$. A multiple solutions state is thus not expected for the Earth. 
One can ask the same question for the Moon. Based on \cite{touma1994evolution}  and \cite{dwyerNature}, considering a Earth-Moon distance varying between its current value and half of this value, and assuming  a constant flattening equal to its current value $e\approx -2.5 \cdot 10^{-5}$, we obtain that $P$ remains of the order $P \approx -10^{-3}$, which is smaller than $P_{res}^{inv} \approx e \approx  -2.5 \cdot 10^{-5}$. Contrary to the Earth, the Moon has thus a Poincar\'e number which is too small for multiple solutions, but one can notice that planetary typical values do not allow to discard the possibility of multiple solutions on simple orders of magnitude arguments. 

Note that, because of its synchronized state, the Moon is rather a precessing triaxial ellipsoid than a spheroid rotating along its symmetry axis \cite{Noir2013}. The solutions are then time-periodic, and one can thus wonder if such multiple solutions can still exist in this case. Figure \ref{fig:kida}b shows that the model proposed  by \cite{Noir2013}, and solved by the script FLIPPER (supplementary material), allows these multiple solutions when the reduced viscous term (\ref{addhocvisc}) is used. As already noticed by \cite{Noir2013}, the solutions in the triaxial ellipsoid remain close from their analogs in the spheroids. Having checked that these multiple solutions also exist using the other viscous terms (\ref{eq:GenVisc}) and (\ref{eqNoSUP}), we thus believe that our estimates for the Moon are quite accurate. A more detailed study of the triaxial ellipsoid case is beyond the scope of this paper.

\section{Conclusion}

In this work, we investigate the ranges of parameters allowing multiple solutions for the flow forced by precession in a spheroid. To do so, we first solve the equations very efficiently, with the dedicated script FLIPPER, provided as a supplementary material. Then, we obtain various analytical results on the solutions. For instance, we obtain analytical estimates of the ranges of parameters allowing these multiple solutions, and these analytical results are successfully compared with numerical solutions of the equations. Finally, the stability of the solutions is analytically obtained, extending the results of \cite{Noir2003} in a quantitative manner. This dynamical model approach also allows an accurate description of the bifurcation of the unstable flow solution.

Naturally, it would be interesting to investigate exprimentally or numerically these co-existent solutions, which have not been observed yet. However, the required values of the Ekman number are quite small, preventing an easy use of local methods. Moreover, usual spherical harmonics codes can only deal with a spherical geometry, where a unique solution is always expected. In order to investigate  this issue, we plan to develop a spectral method designed to deal with spheroids.

\ack{
The author is grateful to J. Noir for enlightening discussions around this topic. }
\appendix

\section{Geometrical interpretation, constraints on the solutions} \label{sec:geom}

In the set of equations (\ref{eq:NoSpinUp})-(\ref{eq:Busse3}), for a given $\Omega_z$, each equation admits solutions in the plane $(\Omega_x,\Omega_y)$. In this plane, the solution of the complete system is then given by the intersection points of these different solutions locations.

The so-called no spin-up equation (\ref{eq:NoSpinUp}) describes a sphere of center $(0,0,1/2)$ and radius $1/2$. For a given $\Omega_z$, solutions are thus on circles of center $(0,0)$, and radius 
\begin{eqnarray}
r=\sqrt{\Omega_z(1-\Omega_z)}=\sqrt{1/4-(\Omega_z-1/2)^2}.
\end{eqnarray}
 Since, $r^2=1/4-(\Omega_z-1/2)^2 \geq 0$, we see that $\Omega_z \in [0; 1]$, and $\Omega_x,\, \Omega_y \in [0; 1/2]$. In the following, we are then considering geometrical constraints for a given $\Omega_z$, i.e. in the plane $(\Omega_x,\Omega_y)$.

Equation (\ref{eq:Busse2}) can be rewritten as $\Omega_y=f_1 \Omega_x+g_1$, describing a ligne. This line goes through the origin ($P_y=0$), which leads to two points of intersection with the circle. The distance $d$ of this line from the circles centers described by equation (\ref{eq:NoSpinUp}), i.e. to the origin, is given by $d^2=g_1^2/(1+f_1^2)$. A solution of the system of equations (\ref{eq:NoSpinUp}-\ref{eq:Busse2}) is thus only possible if an intersection point exists, i.e. if $d^2 \leq r^2  \Leftrightarrow  g_1^2 \leq \Omega_z (1-\Omega_z)(1+f_1^2)$. For $x=1-\Omega_z \ll 1$, the expansion of $\Omega_z (1-\Omega_z)(1+f_1^2)/g_1^2$ leads to $\Omega_z \leq 1$, using $E \ll 1$. This is not a supplementary constraint on $\Omega_z$.

Equation (\ref{eq:Busse3}) can be rewritten as $\Omega_y =f_2 \Omega_x+g_2$, which describes a line. This line is horizontal when $B=0$, since we then have $\Omega_y=g_2$ . This leads to the constraint $ g_2^2 \leq \Omega_z (1-\Omega_z)(1+f_2^2)$. For $x=1-\Omega_z \ll 1$, the expansion of $ g_2^2 /[\Omega_z (1-\Omega_z)(1+f_2^2)] $ leads to $\Omega_z \geq 1- P_x^2/(\lambda_r^2 E)$, which is trivial since $E \ll 1$. This is thus always verified in this limit.

The two lines are crossing when $\Omega_y=f_1 \Omega_x+g_1=f_2 \Omega_x+g_2 $, i.e. $ \Omega_x=(g_2-g_1)/(f_1-f_2)$. We thus have $| \Omega_x| \leq r $, i.e.
\begin{eqnarray}
\frac{g_2-g_1}{f_1-f_2} \leq \sqrt{\Omega_z(\Omega_z-1)}.
\end{eqnarray}
For $x=1-\Omega_z \ll 1$, the expansion leads to
\begin{eqnarray}
\Omega_z \geq 1- \frac{P_x^2}{(P_z+\eta_3 - \eta_2+\lambda_i \sqrt{E})^2},
\end{eqnarray}

\section{Reduced model: calculation details} \label{sec:calc}
 Focusing on stationary solutions of equations (\ref{eq_axi_P_x})-(\ref{eq_axi_P_z}), these equations can be recast in a polynomial of degree $3$ for the unknown $\Omega_z$. Indeed, equation (\ref{eq_axi_P_z}) gives
  \begin{eqnarray}
\Omega_y=\frac{(\Omega_z-1)\lambda_r \sqrt{E}}{P_x(1+e)}, \label{eq:Wy}
\end{eqnarray}
which can be replaced in equation (\ref{eq_axi_P_x}), leading to
  \begin{eqnarray}
\Omega_x= \frac{[e(\Omega_z+P_z)-P_z)](\Omega_z-1)}{P_x(1+e)}. \label{eq:Wx}
\end{eqnarray}
 Finally, using these two expressions, (\ref{eq_axi_P_y}) can be written as
 \begin{eqnarray}
e^2 \, \Omega_z^3&+&e (2 e c_o P-2 c_o P-e)\, \Omega_z^2+(P^2 s_i^2+e^2 c_o^2 P^2+P^2 s_i^2 e-2 e c_o^2 P^2\\ &-&2 e^2 c_o P+K^2+2 e c_o P+c_o^2 P^2)\, \Omega_z-K^2+[e(2-e)-1] c_o^2 P^2\, \\ & &  = 0, \label{eq:appEQ}
\end{eqnarray}
where $c_o=\cos \alpha$, $s_i=\sin \alpha$ and $K=\lambda_r \sqrt{E}$. One can first notice that, for the sphere ($e=0$), this equation reduces to a linear polynomial, which gives the solution (\ref{eq:sphereX})-(\ref{eq:sphereZ}).

We are interested by the number of solutions of equation (\ref{eq:appEQ}), which implies to study the sign of the discriminant $\Delta$ of this cubic equation. It turns out that the discriminant $\Delta$ is given by an equation of degree $3$ in $k=\lambda_r^2 E$, which allows to obtain explicitly the roots $k_1$, $k_2$, and $k_3$, of $\Delta$.  An expansion for $P \ll 1$ gives at order $\mathcal{O}(P^5)$
\begin{eqnarray} \label{k1}
k_1=\frac{1-e^2}{e} s_i^2 c_o P^3+\frac{1+e}{4e^2}s_i^2[(4 c_o^2(1+e^2)+s_i^2(1+e)-8e c_o^2] P^4,
\end{eqnarray}
and, at order $\mathcal{O}(P^3)$,
\begin{eqnarray}
 \mkern-36mu  k_2 &=&-e^2+2e[c_o(1-e)-\vartheta]P+[\vartheta_2 c_o^2+ c_o (1-e) \vartheta -\frac{3}{2} s_i^2(1+e) ] P^2, \\
 \mkern-36mu  k_3 &=& -e^2+2e[c_o(1-e)+\vartheta]P+[\vartheta_2 c_o^2- c_o (1-e) \vartheta -\frac{3}{2} s_i^2(1+e) ] P^2,
\end{eqnarray}  
where $\vartheta=s_i \sqrt{2(1+e)}$, and $\vartheta_2= 2-e^2-1$. The root $k_1$ corresponds actually to $P_s$ and $P_{s2}$ (equation \ref{eq:Es}), and $k_1$ is thus important for the caracterization of the multiple soution zone. A more accurate expansion of this important root has been obtained for $\alpha=\pi/2$, which is given by equation (\ref{eq:Es90}).

We consider now  the inviscid limit $E=0$, which gives $k_1=0$. At the order of equation (\ref{eq:Es}), $k_1=0$ for two different values of $P$. The first one is $P=0$, which corresponds to the inviscid limit $P_{s}^{inv}=0$ of $P_{s}$. The second one is given by equation (\ref{Pk10}), which is the inviscid limit $P_{s2}^{inv}$ of $P_{s2}$.

Focusing on the inviscid limit $E=k/\lambda_r^2=0$, the discriminant $\Delta$ reduces to $D_0=\Delta(k=0)$. Noting that $P=0$ is a solution, we consider $D_0/P^3$, which is an algebraic equation of degree $3$ in $P$. Since $\Delta$ is the discriminant of equation (\ref{eq:appEQ}), the roots of $D_0$ naturally correspond to the multiple solutions zones boundaries in the inviscid limit. Thus, the roots of $D_0$ will allow to obtain $P_{res}^{inv}$, $P_{res2}^{inv}$, $P_{s}^{inv}$, and $P_{s2}^{inv}$. The solution $P=0$ of $D_0$ is naturally the  inviscid limit $P_{s2}^{inv}=0$ of $P_{s2}$ (which has been seen above). The three remaining roots, which are the analytically known roots of $D_0/P^3$, naturally correspond to $P_{res}^{inv}$, $P_{res2}^{inv}$ (equations \ref{eq:Pres} and \ref{eq:Pres2inv}), and $P_{s2}^{inv}$, given by the following expansion
\begin{eqnarray}
P_{s2}^{inv} \approx \frac{4e(e-1)}{1+e} x+\frac{2e(1-e)(24e^2-53e+19)}{3(1+e)^2} x^3+\mathcal{O}(x^5), \label{Pres3}
\end{eqnarray}
where $\alpha=\pi/2-x \ll 1$. Note that these expressions of $P_{s2}$ and $P_{res2}^{inv}$ are not real when $\alpha$ is lower than a certain value $\alpha_{lim}$, determined below, and their expansions around $\alpha=0$ are thus not relevant. An expansion of (\ref{Pk10}) around $x=0$ allows to recover exactly the expression (\ref{Pres3}) at order $4$ (but the term in $x^5$ differs in the two expansions). In the invscid limit, $P_{s2}^{inv}$ is thus given accurately by (\ref{Pk10}) for $P \ll 1$ and arbitrary $\alpha$, and by (\ref{Pres3}) for arbitrary $P$ but $\pi/2-\alpha \ll 1$. 

The sign of $\Delta=0$ gives the number of solution for $\Omega_z$, i.e. the zones where mutiple solutions are possible. As shown in figure \ref{fig:mult2}a, two zones of multiple solutions can exist in the plane $(E,P)$. In the inviscid limit, the number of zones is thus directy given by the number of solution of $D_0=0$, i.e. by the sign of the discriminant $\Delta_0$ of $D_0/P^3$. This discriminant is given by
\begin{eqnarray}
\Delta_0=\frac{s_i^4 e^6(1+e)^2}{432} \frac{[(27e^2-53e+28) c_o^2-1-e]^3}{[e(e-3)c_o^2-e-1]^8},
\end{eqnarray}
which is equal to $\Delta_0=0$ for $\alpha=0$ or
\begin{eqnarray}
\cos \alpha =\pm \sqrt{\frac{1+e}{27e^2-53e+28}},
\end{eqnarray}
which naturally gives the inviscid limit $\alpha_{lim}^{inv}$ of $\alpha_{lim}$ (equation \ref{eq:angleLIM}).

By definition,  $E_{max}$ is the intersection point of $P_s$  and $P_{res}$, whereas  $E_{max2}$ is the intersection point of $P_{s2}$  and $P_{res2}$. One can thus use our previous estimates to calculate them. To do so, we replace $P$ in $\Delta=0$ (which defines $P_s$ and $P_{s2}$), by the inviscid estimates $P_{res}^{inv}$ and $P_{res2}^{inv}$ of, respectively, $P_{res}$ and $P_{res2}$  (equations \ref{eq:Pres} and \ref{eq:Pres2inv}). This gives respectively $E_{max}$  and $E_{max2}$ (from the roots $k=\lambda_r^2 E$ of $\Delta$).\\

\bibliographystyle{plain}
\bibliography{reference_precession}

\end{document}